\documentclass[12pt]{article}

\newcommand{\ket}[1]{ \mbox{$|\, #1\,\rangle$}}

\def\sumint{\hbox{\small$\Sigma$}\kern-0.73em\int\kern.1em}

\renewcommand{\Re}{\operatorname{Re}}

\usepackage{lineno}
\usepackage{authblk}
\usepackage{scicite}
\usepackage{times}
\usepackage{graphicx}
\usepackage{xcolor}
\usepackage{array,multirow,graphicx}
\usepackage{multirow,hhline,graphicx,array}
\usepackage{rotating}
\usepackage{caption}
\usepackage{soul}
\usepackage{hyperref}
\usepackage{mathtools}
\usepackage[superscript,biblabel,nomove]{cite}
\usepackage{amsmath}
\usepackage{a4wide}
\usepackage{bm}
\usepackage{babel}
\usepackage{fontenc}
\usepackage{tikz}
\usepackage{subfigure}
\usepackage{braket}
\usepackage{microtype}

\newcolumntype{M}[1]{>{\centering\arraybackslash}m{#1}}

\usepackage{url,hyperref}
\hypersetup{colorlinks=true, linkcolor=blue, citecolor=blue, urlcolor=blue}

\usepackage[resetlabels,labeled]{multibib}
\newcites{supp}{Supplementary References}

\def\equationautorefname#1#2\null{Eq.#1(#2\null)}

\newenvironment{sciabstract}{%
\begin{quote} \bf}
{\end{quote}}
\author[1,*]{Sooraj Rajendran}\author[2,*]{Miguel Benito de Lama}\author[1,3,*]{Praveen Kumar Maroju}\author[4,5]{Michele Di Fraia}\author[5]{Oksana Plekan}\author[1,3]{David Busto}\author[1]{Ioannis Makos}\author[1]{Marvin Schmoll}
\author[5,6]{Luca Giannessi}\author[5]{Enrico Allaria}\author[5]{Primo\v{z} Rebernik Ribi\v{c}}\author[5,7]{Giovanni De Ninno}%\author[14]{Jianxiong Li}
\author[5]{Alexander Demidovich}\author[5]{Miltcho Danailov}
\author[4,5]{Marco Zangrando}
\author[8]{Kenneth J. Schafer}
\author[9]{Richard J. Squibb}\author[9]{Raimund Feifel}

\author[10]{Tam\'{a}s Csizmadia}

\author[11]{Fabio Frassetto}
\author[11]{Luca Poletto}
\author[5,12]{Kevin C. Prince}\author[3]{Johan Mauritsson}\author[5]{Carlo Callegari}\author[13,14]{Johannes Feist$\dagger$}\author[2,14,15]{Alicia Palacios$\dagger$}\author[1,16]{Giuseppe Sansone$\dagger$}
\affil[1]{Institute of Physics, University of Freiburg, Stefan Meier Strasse 19, 79104 Freiburg, Germany.}
\affil[2]{Departamento de Química, Módulo 13, Universidad Autónoma de Madrid (UAM), 28049 Madrid, Spain}

\affil[3]{Department of Physics, Lund University, PO Box 118, SE-221 00 Lund, Sweden.}
\affil[4]{CNR - Istituto Officina dei Materiali (IOM), 34149 Basovizza, Trieste, Italy.}
\affil[5]{Elettra-Sincrotrone Trieste S.C.p.A., 34149 Basovizza, Trieste, Italy.}
\affil[6]{INFN Laboratori Nazionali di Frascati, Via E. Fermi 54, 00044 Frascati (Rome) Italy.}
\affil[7]{Laboratory of Quantum Optics, University of Nova Gorica, 5001 Nova Gorica, Slovenia.}
\affil[8]{Department of Physics and Astronomy Louisiana State University Baton Rouge, Louisiana 70803-4001, USA.}
\affil[9]{Department of Physics, University of Gothenburg, Origov\"{a}gen 6B, 412 96 Gothenburg, Sweden.}
\affil[10]{ELI ALPS, ELI-HU Non-Profit Ltd., Wolfgang Sandner utca 3., H-6728 Szeged, Hungary.}
\affil[11]{Istituto di Fotonica e Nanotecnologie, CNR, 35131 Padova, Italy.}
\affil[12]{Charles University, Faculty of Mathematics and Physics, Department of Surface and Plasma Science, V Holešovičkách 2, Prague, 18000, Czech Republic.}
\affil[13]{Departamento de Física Teórica de la Materia Condensada, UAM, 28049 Madrid, Spain.}
\affil[14]{Condensed Matter Physics Center (IFIMAC), UAM, 28049 Madrid, Spain.}
\affil[15]{Institute for Advanced Research in Chemical Sciences (IAdChem), UAM, 28049 Madrid, Spain.}
\affil[16]{Freiburg Institute for Advanced Studies (FRIAS), University of Freiburg, Albertstraße 19, 79104 Freiburg, Germany}
\affil[*]{contributed equally to this work}

\title{Breakdown of the isotropic asymptotic approximation in two-colour photoionisation.}%Single-shot sideband correlation analysis for attosecond waveform synthesis}

\date{}

\begin{document}

% Double-space the manuscript.

\baselineskip24pt

% Make the title.

\maketitle

% Place your abstract within the special {sciabstract} environment.

\begin{sciabstract}
The Wigner delay is defined as the energy derivative of the scattering phase of a particle in a given potential, unveiling the time taken (or gained) due to the interaction. 
The characterisation of this delay plays a central role in attosecond science, where the time resolution allows to gain information on the time interval required for a photoelectron to be emitted into the continuum after the absorption of a single photon. Attosecond interferometric techniques, based on two-colour (extreme ultraviolet and near-infrared) photoionisation schemes, cannot provide a direct measurement of the Wigner delay, because the low-frequency photon contributes with an additional delay, which is imprinted on the outgoing photoelectron. The isolation of the Wigner delay is usually achieved by appealing to the asymptotic approximation, which assumes that the two-photon delay is separable into a Wigner and a near-infrared-induced phase and provides a universal analytical expression for the latter.

In this study, we introduce a self-referencing approach based on the implementation of non-consecutive extreme ultraviolet harmonics, in order to test the validity of the asymptotic approximation. We demonstrate its breakdown by observing a deviation of a few tens of milliradians (corresponding to a few attoseconds) between its predictions and the experimentally measured phases of the sideband oscillations generated in our scheme, in agreement with full-dimensional simulations.

\end{sciabstract}
\subsection*{Introduction}
The measurement of attosecond time delays in photoionisation is an extremely active research field with the main goal of gaining information about the electronic structure of simple and complex targets and their dynamics with unprecedented temporal resolution~\cite{pazourekAttosecondChronoscopyPhotoemission2015}. Experimental and theoretical investigations have stimulated extensive debate regarding the interpretation of the physical significance~\cite{maquetAttosecondDelaysPhotoionization2014} and the correct definition of such delays~\cite{feticWignerTimeDelay2024}. The principal technique used for the characterisation of attosecond time delays is based on the two-colour photoionisation process triggered by isolated~\cite{NATURE-Cavalieri-2007,schultzeDelayPhotoemission2010} and trains~\cite{klunderProbingSinglePhotonIonization2011,isingerPhotoionizationTimeFrequency2017} of attosecond pulses in the extreme ultraviolet (XUV), which emit a photoelectron into the continuum in the presence of a near-infrared (NIR) field. In the scheme with attosecond pulse trains, the presence of the latter determines the appearance of sidebands of the main photoelectron peaks~\cite{gloverObservationLaserAssisted1996,schinsCrosscorrelationMeasurementsFemtosecond1996}, whose intensities change as a function of the relative delay $\tau$ between the XUV and NIR radiation~\cite{veniardPhaseDependenceN+1color1996}. From the phase of the oscillations of the sidebands, the phase of the harmonic radiation or information on the phases imparted by the photoionisation process can be gained. This experimental technique is usually referred to as Reconstruction of Attosecond Beating by Interference of Two-photon transitions (RABBIT), and it has been widely used for attosecond metrology~\cite{paulObservationTrainAttosecond2001,mullerReconstructionAttosecondHarmonic2002,hanAttosecondMetrologyCircular2023} and spectroscopy in atoms~\cite{klunderProbingSinglePhotonIonization2011}, molecules~\cite{haesslerPhaseresolvedAttosecondNearthreshold2009,huppertAttosecondDelaysMolecular2016} and solids~\cite{inzaniAttosecondElectronDynamics2025}. In particular, the implementation of this approach has evidenced fundamental aspects of the photoionisation process including the role played by electronic correlation~\cite{sansoneElectronCorrelationReal2012, koturSpectralPhaseMeasurement2016, grusonAttosecondDynamicsFano2016,isingerPhotoionizationTimeFrequency2017}, the contribution of shape resonances~\cite{nandiAttosecondTimingElectron2020, ahmadiAttosecondPhotoionisationTime2022}, the influence of NIR-induced transitions in the ionised target~\cite{makosEntanglementPhotoionisationReveals2025}, and the effect of the lifetimes of the final states~\cite{taoDirectTimedomainObservation2016}.

In the two-colour photoionisation of atoms, the phase imprinted by the combined action of the XUV and NIR fields can be decomposed in two terms. The first one is the scattering phase of the photoelectron after one-photon XUV photoionisation and corresponds to the Wigner phase ($\phi_{\mathrm{wig}}$). Its energy derivative is the Wigner delay $\tau_{\mathrm{wig}}$~\cite{wignerLowerLimitEnergy1955}, which is a very sensitive probe of the electronic structure of the target system. The second one, usually denoted as continuum-continuum phase ($\phi_{cc}$)~\cite{JPB-Dahlstrom-2012,dahlstromTheoryAttosecondDelays2013,pazourekAttosecondChronoscopyPhotoemission2015}, and its corresponding delay $\tau_{cc}$, is due to the interaction between the photoelectron wave packet leaving the target and the NIR field. The decomposition relies on the asymptotic approximation that takes its name because the wave functions in the electronic continuum are approximated with their asymptotic form, i.e. far away from the parent ion. In its original formulation~\cite{JPB-Dahlstrom-2012}, the approximation is equivalent to neglecting the centrifugal potential term, returning results that are independent of the angular momentum quantum number of the electronic wave function. Furthermore, the approach also entails additional approximations such as considering an initial $s$-state, extending the lower limit of a principal-value integral, while the continuum spectra begins at zero, or neglecting a counter-rotating integrand term oscillating with a sum of momenta relative to a co-rotating one with their difference. Hereafter, we will refer to this approach as isotropic asymptotic approximation to point out its \textit{l}-independent character. Under these simplifications, the continuum-continuum phase acquires a universal character~\cite{bustoAttosecondDynamicsNonresonant2024}, independent of the angular momentum of the final state in the continuum, and is described by an analytic expression~\cite{JPB-Dahlstrom-2012}.
Therefore, the  \textcolor{black}{isotropic} asymptotic approximation enables the extraction of the Wigner phase from the experimentally measured multiphoton photoionization phases by simply subtracting the continuum-continuum term~\cite{kheifetsWignerTimeDelay2023a}. Moreover, the latter is assumed independent on the target and simply understood as a measurement-induced additional phase. 
The \textcolor{black}{isotropic asymptotic} approximation has also demonstrated to give a reliable estimation, within the experimental error bars, of the continuum-continuum phase in experiments where it was isolated by combining co- and counter-rotating XUV and NIR circularly polarized fields~\cite{hanSeparationPhotoionizationMeasurementinduced}. 

More refined approximations, which include $l$-dependent effects have been demonstrated in the last years, by replacing the approximate wave function for the final electronic state with exact solutions of the Coulomb equations~\cite{bollAnalyticalModelAttosecond2022, bollTwocolorPolarizationControl2023}, and by considering exact solutions for the final as well as intermediate~states~\cite{bendaAngularMomentumDependence2025,jiAnalyticalExpressionContinuum2024}. Furthermore, by using a WKB approach, the original isotropic asymptotic approximation was extended to include the effect of the angular momentum of the final state on the electronic wave function~\cite{berkaneProbingWignerTime2024}.

For sufficiently high intensities, the interaction of the NIR field with an outgoing photoelectron wave packet can lead to the exchange of several NIR photons, leading to the formation of high-order sidebands~\cite{radcliffeExperimentTwocolorPhotoionization2007,radcliffeSingleshotCharacterizationIndependent2007}. In the case of multiple NIR-photon transitions, it has been shown that, within the isotropic asymptotic approximation, the overall phase in photoionisation can still be represented as the sum of a Wigner and a continuum-continuum term. Furthermore the latter can be decomposed into the sum of the continuum-continuum phases of the intermediate steps. Each of these terms corresponds to the exchange (absorption or emission) of a single NIR photon~\cite{bhartiDecompositionTransitionPhase2021, bhartiMultisidebandInterferenceStructures2024, bertolinoMultiphotonInteractionPhase2021}: $\phi_{cc}\approx \sum_{i=1}^{n}\phi_{cc}^{i}$, where $n$ indicates the number of exchanged NIR photons. The phase accumulated by the photoelectron wave packet in the scenario characterised by multi-NIR photon transitions has been recently discussed~\cite{bertolinoMultiphotonInteractionPhase2021}.

The isotropic asymptotic approximation fails to correctly interpret the angular-dependence of attosecond time delays measured in the photoionisation of atoms~\cite{heuserAngularDependencePhotoemission2016b, OPT-Fuchs-2020}. Within this approach, the ratio of the transition matrix elements for the absorption and emission of NIR photons is, in fact, independent of the angular momentum of the final state, in direct contradiction to the Fano propensity rule~\cite{bustoFanosPropensityRule2019}. However, despite the large number of works based on the isotropic asymptotic approximation, the quantitative limits of its estimation of the phase introduced by the NIR pulse has not yet been the subject of a benchmark experiment. It is important to point out that the isotropic asymptotic approximation is expected to be accurate within a few tens or few attoseconds, depending on the kinetic energy of the emitted photoelectron. This calls for an approach that can determine delays (or, equivalently, phase differences) with comparable accuracy in order to verify its validity. This is a challenging request for experiments, as it requires the isolation of the continuum-continuum contribution from other phases occurring in the photoionisation process, such as the Wigner phase and the phase of the XUV radiation.

\begin{figure}[!htbp]
\centering \resizebox{1.0\hsize}{!}{\includegraphics{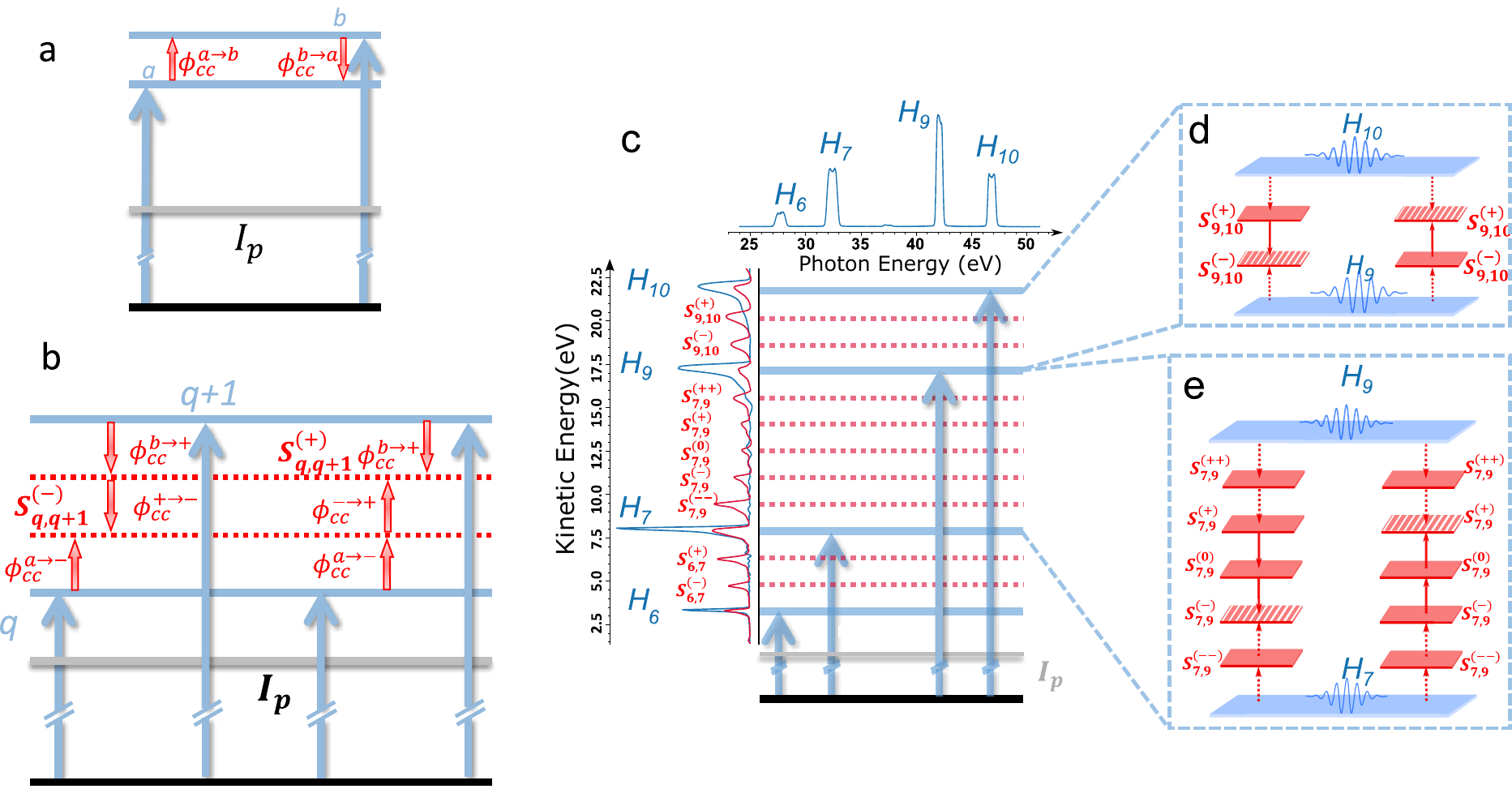}}
\caption{\textbf{Energy scheme and self-refencing approach for the investigation of the breakdown of the isotropic asymptotic approximation in two-colour photoionization.} a) Schematic view of the energy levels of a two-colour photoionisation scheme leading to the population of the two energy levels $a$ and $b$ through the exchange (absorption and emission) of a single NIR photon. $I_p$ indicates the ionisation potential of the target atom. b) Schematic view of the energy levels of a self-referencing two-colour photoionisation scheme based on the measurement of two sidebands $S^{(+)}_{q,q+1}$ and $S^{(-)}_{q,q+1}$ between the main photoelectron peaks generated by the absorption of one XUV photon from the harmonics $q$ and $q+1$ of a driving field with angular frequency $3\omega$. c) XUV photon spectra (upper line) consisting of the four non-consecutive harmonics $H_6$, $H_7$, $H_9$ and $H_{10}$. Photoelectron spectra (left hand side) generated in helium by single-photon ionisation using the XUV harmonics with (red spectrum) and without (blue spectrum) the NIR field. The energy levels reached in the single and two-photon ionisation processes are schematically presented. d) Photoionisation pathways leading to the population of the sidebands $S^{(\pm)}_{9,10}$ between the consecutive harmonics $H_9$ and $H_{10}$. The red arrows with solid lines indicate the exchange of NIR photons that differs in the population between the sidebands $S^{(-)}_{9,10}$ and $S^{(+)}_{9,10}$. e) Photoionisation pathways leading to the population of the sidebands $S^{(\pm)}_{7,9}$ between the nonconsecutive harmonics $H_7$ and $H_{9}$. Similarly to panel d, the red arrows with solid lines indicate the exchange of two NIR photons that differs in the population between the sidebands $S^{(+)}_{7,9}$ and $S^{(-)}_{7,9}$.}
\label{Fig1}
\end{figure}

In this work, we introduce a self-referencing experimental approach that does not require the characterisation of any additional phase or delay contributions, allowing the isotropic asymptotic approximation to be validated with a resolution of a few attoseconds.
The idea behind our approach is shown in Fig.~\ref{Fig1}a. It is based on the direct implication of the isotropic asymptotic approximation that the phases acquired in the absorption ($\phi_{cc}^{a\rightarrow b}$) and emission ($\phi_{cc}^{b\rightarrow a}$) between two energy levels $a$ and $b$ of the continuum are exactly opposite, i.e., their sum is zero~\cite{JPB-Dahlstrom-2012}:
\begin{equation}\label{phi_cc_sum}
\phi_{cc}^{a\rightarrow b}+\phi_{cc}^{b\rightarrow a}=0.
\end{equation}
Any deviation in the sum of the two continuum-continuum phases from this value would indicate a violation or breakdown of the approximation.

Despite the simplicity of the idea, this scheme cannot be easily realised in an experiment. For a photoionisation process triggered by a single XUV frequency in the presence of the NIR field, the continuum-continuum phase must be isolated independently for the absorption and emission path, by measuring the total photoionisation phase of the photoelectron wave packet emitted in the two-colour process. Furthermore, the Wigner phase must be measured in the XUV-only photoionisation case and substracted from the two-colour case to isolate the continuum-continuum phase. Finally, this approach requires the previous characterisation of the phases of the two XUV fields ($\varphi_a$ and $\varphi_b$) leading to the population of the two continuum states $a$ and $b$.

Alternatively, an interferometric technique could be used to extract the phase difference between the two levels. This technique could be based on the use of two coherent XUV fields and a synchronised NIR field with an energy matching the energy difference between the two XUV fields, as well as a variable delay $\tau$. However, the photoelectron signal measured in this approach would not be background-free, resulting in a complex reconstruction procedure of the phases of the photoelectron peak oscillations that generally would depend on the intensity of the two XUV pulses~\cite{loriotHighHarmonicGeneration2o2020a, laurentAttosecondControlOrbital2012}.

To overcome these limitations, we present a method that provides direct access to the sum of the continuum-continuum phase between two energy levels, as illustrated in Fig.~\ref{Fig1}b.
Two consecutive, XUV, coherent harmonics ($q$ and $q+1$ with phases $\varphi_q$ and $\varphi_{q+1}$) of a fundamental radiation with an angular frequency $3\omega$ are used to photoionise an atomic target in the presence of a synchronised replica of the radiation at frequency $\omega$ with a variable delay $\tau$. The presence of the radiation with frequency $\omega$ determines the generation of two sidebands ($S^{(-)}_{q,q+1}$ and $S^{(+)}_{q,q+1}$) of the main photoelectron peaks. Population of the sideband $S^{(-)}_{q,q+1}$ ($S^{(+)}_{q,q+1}$) occurs via two coherent paths, characterised by the absorption of one (two) and emission of two (one) NIR photons of frequency $\omega$. Due to the coherent superpositions of the two paths, the signal of the sidebands $S^{(-)}_{q,q+1}$ and $S^{(+)}_{q,q+1}$ is expected to oscillate with phases $\phi^{(-)}$ and $\phi^{(+)}$, respectively. Because the final states along the two pathways exhibit different parity, in the experiment the photoelectron signal is integrated only over the half-space along the common polarisation direction of the XUV and NIR fields to enable the observation of the yield oscillations.

Within the isotropic asymptotic approximation, the phases of the two sidebands can be decomposed into different terms, each of them corresponding to the phase associated with the exchange of a single photon from the two-colour (XUV-NIR) field, according to the expressions~\cite{bhartiDecompositionTransitionPhase2021}:
\begin{eqnarray}% \nonumber % Remove numbering (before each equation)
\label{Eqdecomp}
 \phi^{(+)}&=&(\varphi_{q+1}-\varphi_{q})+(\phi_{\mathrm{wig}}^{q+1}-\phi_{\mathrm{wig}}^q)+\phi_{cc}^{q+1\rightarrow +}-[\phi_{cc}^{q\rightarrow -}+\phi_{cc}^{-\rightarrow +}]+3\omega\tau+\pi/2\nonumber\\
 \phi^{(-)}&=&(\varphi_{q+1}-\varphi_{q})+(\phi_{\mathrm{wig}}^{q+1}-\phi_{\mathrm{wig}}^q)+[\phi_{cc}^{q+1\rightarrow +}+\phi_{cc}^{+\rightarrow -}]-\phi_{cc}^{q \rightarrow -}+3\omega\tau-\pi/2,
\end{eqnarray}
where $\phi_{\mathrm{wig}}^{q+1}$ and $\phi_{\mathrm{wig}}^{q}$ are the Wigner phases associated to photoionisation with the harmonics $q+1$ and $q$, and the factors $\pm\pi/2$ arise from the fact that, for each sideband, the two contributing paths exchange a different number of photons.
The difference of the two phases depends only on the continuum-continuum phase according to the relation:
\begin{equation}\label{Eq0}
  \phi^{(+)}-\phi^{(-)}=\phi_{cc}^{+\rightarrow -}+\phi_{cc}^{-\rightarrow +}+\pi
\end{equation}

As a result, when comparing the phase of the oscillations of the sidebands $S^{(+)}_{q,q+1}$ and $S^{(-)}_{q,q+1}$, all phase terms that differ from the continuum-continuum phases between the two energy levels vanish, providing access to the sum of the continuum-continuum phases. The latter is zero within the isotropic asymptotic approximation, leading to sidebands that are in perfect phase opposition. Any deviation of the relative phase from $\pi$ would imply a violation of the isotropic asymptotic approximation in photoionisation. 

We observe that without the isotropic asymptotic approximation, it is not possible to associate a single phase to each intermediate step characterised by the exchange of a NIR photon, as expressed in Eq.~\ref{Eqdecomp}. Indeed, since the overall phases depend on the angular momenta of the intermediate and final states, many different pathways contribute to each sideband. In this picture, the overall phases are then not simply the sum of phases for each intermediate steps, but the phase of the sum of complex amplitudes connecting the initial and final states, through all possible combinations of intermediate steps.

We point out that our approach is not sensitive to the value of the continuum-continuum phase and thus to its deviation from the corresponding value extracted from TDSE simulations. The scheme presented in this work relies on a particular attosecond interferometry configuration in which photoelectron wave packets, contributed by paths leading to the same final energy, interfere. Consequently, our scheme is sensitive to the combination (sum) of the continuum-continuum phases appearing in the phases accumulated along the two paths.

Our self-referencing approach was realised at the seeded free-electron laser (FEL) FERMI~\cite{allariaHighlyCoherentStable2012} (see Fig.~\ref{Fig1}c) by using two consecutive harmonics of the seed laser with frequency $3\omega$ ($H_9$ and $H_{10}$, as shown in Fig.~\ref{Fig1}c) obtained as frequency triplication of a NIR field with a wavelength of 798~nm. Using a replica of the same NIR pulse in the XUV-induced photoionisation process generates two sidebands $S^{(\pm)}_{9,10}$ (Fig.~\ref{Fig1}d) between the two harmonics. However, at FERMI, the NIR field and the XUV field have a measured delay jitter of about $\pm3$~fs, thus preventing direct observation of the sideband oscillations by controlling the relative delay between the two fields~\cite{danailovJitterfreePumpprobeMeasurements2014}. To overcome this limitation, we implemented the attosecond timing tool demonstrated in Ref.~\citenum{marojuAttosecondCoherentControl2023}, using two additional harmonics ($H_6$ and $H_7$), thus creating a comb of non-consecutive harmonics~\cite{marojuAttosecondTemporalStructure2025}.

The use of non-consecutive harmonics has three important, additional advantages. Firstly, as our self-referencing approach requires a NIR intensity of a few $10^{12}$~W/cm$^2$ to observe two-NIR photon transitions, the gap between the two groups of harmonics significantly reduces the contribution of multi-NIR-photon-transition paths connecting the two groups of harmonics. By eliminating these contributions, we can experimentally reproduce the ideal conditions of the scheme presented in Fig.~\ref{Fig1}b. Secondly, by using two groups of non-consecutive harmonics, we can investigate the validity of the asymptotic approximation in two different energy intervals within the same experiment. Thirdly, using non-consecutive harmonics we can extend our investigation also to the sidebands between the non-consecutive harmonics $H_7$ and $H_9$. As shown in Fig.~\ref{Fig1}e, the pairs of sidebands $S^{(+)}_{7,9}$-$S^{(0)}_{7,9}$ and $S^{(0)}_{7,9}$-$S^{(-)}_{7,9}$ exhibit terms oscillating at a frequency of $6\omega$ and differ by the absorption or emission of a single NIR photon. At the same time, the sidebands $S^{(+)}_{7,9}$ and $S^{(-)}_{7,9}$ differ in the absorption or emission of two NIR photons. The analysis of the relative phase of the oscillations of these sidebands provides an additional experimental observable to identify the breakdown of the isotropic asymptotic approximation.

\subsection*{Results}
\subsubsection*{Experiment}
\begin{figure}[!htbp]
\centering \resizebox{1.0\hsize}{!}{\includegraphics{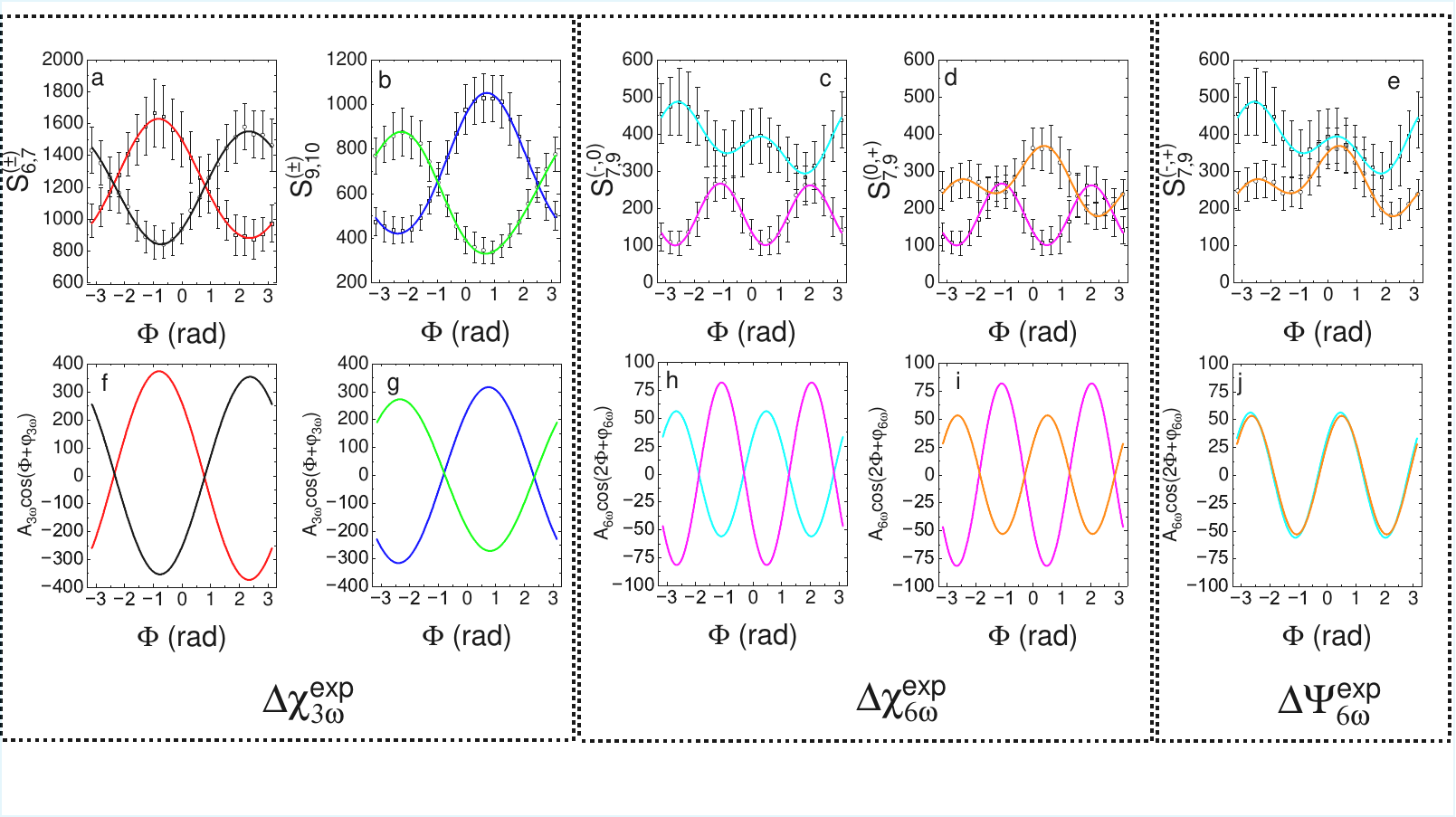}}
\caption{\textbf{Sideband oscillations and determination of the phase differences $\Delta\chi_{3\omega}^{\mathrm{exp}}$, $\Delta\chi_{6\omega}^{\mathrm{exp}}$, and $\Delta\Psi_{6\omega}^{\mathrm{exp}}$.} Experimental oscillations of the sidebands $S^{(\pm)}_{6,7}$ (a), $S^{(\pm)}_{9,10}$ (b), $S^{(-,0)}_{7,9}$ (c), $S^{(0,+)}_{7,9}$ (d), and $S^{(-,+)}_{7,9}$ (e) and corresponding fits according to Eq.~\ref{Eq1}. Component oscillating at $3\omega$ of the fit for the sidebands $S^{(\pm)}_{6,7}$ (f), and $S^{(\pm)}_{9,10}$ (g). Component oscillating at $6\omega$ of the fit for the sidebands $S^{(-,0)}_{7,9}$ (h), $S^{(0,+)}_{7,9}$ (i), and $S^{(-,+)}_{7,9}$ (j). The intensity of the NIR field was $I_2=2.2\times10^{12}$~W/cm$^2$. The second-order phase difference was $\Delta\varphi_{6,7,9,10}=\Delta\varphi_A$. The error bars in the plots were calculated as the standard deviation of the sideband amplitudes for the corresponding phase value $\Phi$.}
\label{Fig2}
\end{figure}

The second-order phase difference of the comb of non-consecutive harmonics presented in Fig.~\ref{Fig1}c was changed by adjusting the phase shifters positioned between each pair of undulators (see Methods section). The harmonic amplitude and energy values are reported in~\ref{Table1}. The sideband oscillations were reconstructed using the attosecond timing tool introduced in Ref.~\citenum{marojuAttosecondCoherentControl2023}. Two settings, indicated in the following as $A$ and $B$, of the second-order harmonic phase difference between two consecutive pairs of harmonics were used, corresponding to $\Delta\varphi_{6,7,9,10}=(\varphi_{10}-\varphi_9)-(\varphi_7-\varphi_6)=\Delta\varphi_{A,B}\approx\pm\pi/2$ (see~\ref{Table2} and~\ref{Table3}). These values of the second-order phase difference indeed deliver the optimal time resolution for the sideband reconstruction~\cite{marojuAttosecondCoherentControl2023}. Measurements in helium were acquired for three intensities of the NIR field: $I_1=1.2\times10^{12}$~W/cm$^2$, $I_2=2.2\times10^{12}$~W/cm$^2$, and $I_3=4.0\times10^{12}$~W/cm$^2$. Additional measurements in neon were acquired for similar intensities (see Supplementary Information).
All sideband oscillations were fitted considering a constant term ($A_0$) and two components oscillating at frequencies $3\omega$ (amplitude $A_{3\omega}$ and phase $\chi^{\mathrm{exp}}_{3\omega}$) and $6\omega$ (amplitude $A_{6\omega}$ and phase $\chi^{\mathrm{exp}}_{6\omega}$):
\begin{equation}\label{Eq1}
  S_{\mathrm{fit}}=A_0+A_{3\omega}\cos(\Phi+\chi^{\mathrm{exp}}_{3\omega})+A_{6\omega}\cos(2\Phi+\chi^{\mathrm{exp}}_{6\omega}),
\end{equation}
where the phase $\Phi$ linearly depends on $3\omega$ and represents the relative phase between the attosecond pulse train and the NIR field (see Supplementary Information).
The reconstructed sideband oscillations for the harmonic configuration $\Delta\varphi_{6,7,9,10}=\Delta\varphi_A\approx-\pi/2$ are presented in Fig.~\ref{Fig2}. The results of the fits are indicated by solid lines in all upper panels (a-e) of Fig.~\ref{Fig2}. The lower panels (f-j) present the evolution of the component relevant for the extraction of the phase difference between the oscillation of the sidebands. In the case of the sidebands $S^{(\pm)}_{6,7}$ (a) and $S^{(\pm)}_{9,10}$ (b) the signal is dominated by the component at $3\omega$. The component of the fit according to Eq.~\ref{Eq1} at $3\omega$ is reported in panels f and g for the two pairs of sidebands.
For the sidebands $S^{(\pm)}_{7,9}$ (panels c-e; cyan for $S^{(-)}_{7,9}$ and orange for $S^{(+)}_{7,9}$) the signal exhibits a complex evolution with comparable contributions of the two components oscillating at 3$\omega$ and 6$\omega$. In contrast, the sideband $S^{(0)}_{7,9}$ (panels c and d; magenta curve) exhibits a dominant oscillation at frequency $6\omega$. Panels h-j present the terms oscillating at frequency $6\omega$ for these three sidebands.

We define the experimental phase difference between the oscillations of these sidebands as follows:
\begin{eqnarray}\label{Eq11}
\Delta\chi^{\mathrm{exp}}_{3\omega}&=&\chi^{\mathrm{exp}}_{3\omega}[S^{(+)}_{q,q+1}]-\chi^{\mathrm{exp}}_{3\omega}[S^{(-)}_{q,q+1}]-\pi\quad q=6,9\nonumber\\
\Delta\chi^{\mathrm{exp};(+0)}_{6\omega}&=&\chi^{\mathrm{exp}}_{6\omega}[S^{(+)}_{7,9}]-\chi^{\mathrm{exp}}_{6\omega}[S^{(0)}_{7,9}]-\pi\quad \nonumber\\
\Delta\chi^{\mathrm{exp};(0-)}_{6\omega}&=&\chi^{\mathrm{exp}}_{6\omega}[S^{(0)}_{7,9}]-\chi^{\mathrm{exp}}_{6\omega}[S^{(-)}_{7,9}]-\pi\quad \nonumber\\
\Delta\Psi^{\mathrm{exp}}_{6\omega}&=&\chi^{\mathrm{exp}}_{6\omega}[S^{(+)}_{7,9}]-\chi^{\mathrm{exp}}_{6\omega}[S^{(-)}_{7,9}],
\end{eqnarray}

As we aim at isolating the sum of the continuum-continuum phase between two levels in the continuum, as expressed in Eq. \ref{Eq1}, the first, second, and third relations contain an additional $\pi$ phase factor. 
 The last term does not contain this additional shift, because the two contributing pathways involve the exchange of the same number of photons.
The phase differences introduced in the first three equations quantify the deviation from the expected value of zero for sidebands oscillating at $3\omega$ and $6\omega$, which differ in the emission or absorption of a single NIR photon ($\Delta\chi^{\mathrm{exp}}_{3\omega}$ and $\Delta\chi^{\mathrm{exp};(+0,0-)}_{6\omega}$). The phase $\Delta\Psi^{\mathrm{exp}}_{6\omega}$ is related to sidebands oscillating at $6\omega$ and differing in terms of the absorption or emission of two NIR photons. 
We distinguish between $\Delta\chi^{\mathrm{exp};(+0)}_{6\omega}$ and $\Delta\chi^{\mathrm{exp};(0-)}_{6\omega}$, because the transitions involve emission and absorption of different angular momentum $l$ in each case. As shown later, the simulated evolution for these phase differences starts to deviate close to the ionisation threshold.

Due to the emission mechanism of the XUV radiation in a seeded FEL also a weak harmonic $H_8$ was generated (see XUV spectrum shown in  Fig.~\ref{Fig1}c for photon energies around 35~eV). This harmonic affects the phase of the terms oscillating at frequencies $3\omega$ and, to a smaller extent, also $6\omega$. The method followed to correct for the shift introduced by the contributions of the weak central harmonic is based on a fitting procedure using the TDSE simulations (see~\ref{Fig0SI}) and is detailed in the Supplementary Information (see~\ref{Fig1SI},~\ref{Fig2SI}, and~\ref{Fig3SI}). The correction depends on the phase of the harmonic $H_8$, whose estimation is presented in~\ref{Table2}. The quantitative analysis of the corrections of the experimental data acquired in helium is presented in~\ref{Table4},~\ref{Table5},~\ref{Table6}, and~\ref{Table7}.

\subsubsection*{Theoretical models}
We solve the full-dimensional time-dependent Schrödinger equation (TDSE) for the helium atom, employing the approach thoroughly described in Refs.~\citenum{feistNonsequentialTwophotonDouble2008,palaciosCrossSectionsShortpulse2008,palaciosTimedependentTreatmentTwophoton2009}. We use a finite-element discrete variable representation (FEDVR) to discretise the radial part of the two-electron wave function and coupled spherical harmonics for the angular component. The electron-laser interaction is treated within the dipole approximation, and time propagation is performed with a short iterative Lanczos algorithm. We extract the ionisation probabilities by directly projecting onto the final scattering states after the external field is turned off.

To mimic the experimental conditions, we perform TDSE simulations with all fields linearly polarised along the same axis, using harmonics 6, 7, 8, 9 and 10 of the seed laser (wavelength $266$~nm), which itself corresponds to the third harmonic of the NIR field (wavelength $798$~nm) used to obtain the RABBIT spectra. All fields have a temporal envelope proportional to $\sin^6(\pi t/T)$ and pulse duration $T=40$~fs. The intensities of the harmonics are set to the values reported in~\ref{Table2}, while the NIR intensity is set to the three experimental values $I_1$, $I_2$, and $I_3$. Instead of performing an explicit scan over delays between harmonics and the NIR field, we use a highly efficient wave function synthesis procedure that allows reconstruction of the full RABBIT spectrogram for arbitrary delays, field phases, and harmonic intensities from a single TDSE simulation carried out for each harmonic and NIR intensity. This procedure, described in detail in the Supplementary Information, exploits the fact that only single-photon absorption from each harmonic contributes to the spectra, and that the sidebands are well-separated in final electron energy. It also gives direct access to the contributions of each pair of harmonics to each sideband, and to the phases of the sideband oscillations without requiring a fitting procedure. We have verified that the results obtained with this procedure are in excellent agreement with those obtained from fitting of the sideband oscillations in explicit delay scans.

We verified that the phase difference obtained with the four non-consecutive harmonics (after correcting for the weak contribution from $H_8$) are equal to those derived considering only two harmonics. Only two harmonics can interfere to produce $6\omega$ oscillations, harmonic 7 and 9, so no additional interfering terms contribute at $6\omega$. For the $3\omega$ oscillations, these arise from interference between harmonics separated by $3\omega$ in energy, namely 6 and 7, or 9 and 10. However, the $3\omega$ oscillating terms between consecutive harmonics are sufficiently separated in energy from the other harmonics to avoid mutual interference. Hence, the upper (lower) pair of harmonics does not interfere with the $3\omega$ oscillations produced by the lower (upper) pair.

Since we are interested in single ionisation and far above the ionisation threshold, a reasonably accurate description can also be obtained using the single-active electron (SAE) approximation, which substantially reduces the computational effort. We use the SAE approximation to perform the simulations in neon, employing a model potential described in Ref.~\citenum{tongEmpiricalFormulaStatic2005}.
We also used different SAE model potentials to test the effect of the short-range potential in the measurement. We use both the hydrogen atom potential and a modified hydrogen potential in which the centrifugal term is removed in order to cleanly separate its contributions. Furthermore, we employ two SAE model potentials that reproduce features of the two-electron He atom: one corresponds to the potential felt by the second electron when the first electron is fixed to the $1s$ level of the He$^+$ ion, while the other is an empirical potential described in Ref.~\citenum{tongEmpiricalFormulaStatic2005} that reproduces the ionisation potential of He accurately.

\subsection*{Discussion}
Figure~\ref{Fig3} presents the comparison of the values measured in helium with the corresponding theoretical simulations for the phase differences $\Delta\chi_{3\omega}$, $\Delta\chi^{(+0,0-)}_{6\omega}$, and  $\Delta\Psi_{6\omega}$. The experimental data were corrected for the effect of the harmonic $H_8$ (see Supplementary Information), while this amplitude was set to zero in the simulations. We show together the phase differences obtained for three different intensities of the NIR field: $I_1$ (green circles), $I_2$ (red squares), and $I_3$ (blue triangles). For visual clarity the data have been shifted by -0.5~eV and +0.5~eV for the lowest and highest intensities, respectively.
Figures~\ref{Fig3}a,b,c and~\ref{Fig3}d,e,f present the data obtained for the second-order phase differences $\Delta\varphi_{6,7,9,10}=\Delta\varphi_A$ and $\Delta\varphi_B$, respectively. The comparison between the experimental data acquired in neon and the corresponding simulations is presented in~\ref{Fig4SI}.

As expected the four simulated phase differences $\Delta\chi_{3\omega}$ (Fig.~\ref{Fig3}a,d), $\Delta\chi^{(+0,0-)}_{6\omega}$ (Fig.~\ref{Fig3}b,e), and $\Delta\Psi_{3\omega}$ (Fig.~\ref{Fig3}c,f) converge to zero for large photoelectron kinetic energies. Within the isotropic asymptotic approximation, these differences are zero regardless of the photoelectron kinetic energy. Furthermore, for the same photoelectron kinetic energy range (for example 10~eV) we observe that the deviation from zero is larger for $\Delta\chi^{(+0,0-)}_{6\omega}$ (approximately 0.05~rad corresponding to 3~as) than for $\Delta\chi_{3\omega}$ (approximately 0.025~rad corresponding to 1.5~as). This is due to the fact that, even though both phase differences differ only for the exchange of a single photon, the former requires the exchange of a greater total number of NIR photons than the latter (see Fig.~\ref{Fig1}), thus enhancing the contributions of terms that are neglected in the isotropic asymptotic approximation. Similarly, we observe that the deviation for the phase difference $\Delta\Psi_{6\omega}$ is generally larger than $\Delta\chi^{(+0,0-)}_{6\omega}$ (approximately 0.1~rad at 10~eV). We attribute this difference to the fact that the former depends on the difference between two pathways that differ in the absorption or emission of two NIR photons, whereas the deviation of the latter from zero is only due to the difference of a single NIR photon (see Fig.~\ref{Fig1}c).

The theoretical predictions for $\Delta\chi_{3\omega}$ (Fig.~\ref{Fig3}a,d) are in reasonable agreement with the experimental results for both phase configurations and for the three intensities. 

For $\Delta\varphi_{6,7,9,10}=\Delta\varphi_A$, the experimental points for $\Delta\chi^{(+0,0-)}_{6\omega}$ (Fig.~\ref{Fig3}b) and $\Delta\Psi_{6\omega}$ (Fig.~\ref{Fig3}c) and the theoretical predictions agree well. Furthermore, it is clear that the error bars diminish as the NIR intensity increases. This is due to the increased contribution of multi-NIR photon transitions, which eventually leads to a better contrast of the $6\omega$ component of the sideband oscillations. In particular, the reduced error bars clearly indicate that the experimental points for the phase difference $\Delta\chi^{(+0,0-)}_{6\omega}$ at the highest intensity are compatible with the TDSE predictions (at the level of one standard deviation $\sigma$), but not with the predictions of the isotropic asymptotic approximation. For the phase difference $\Delta\Psi_{6\omega}$ the experimental points acquired at the two highest intensities are not compatible (within 1.6$\sigma$ and 2.5$\sigma$) with the isotropic asymptotic approximation and compatible within one $\sigma$ with the TDSE simulations. This observation offers a further indication of the breakdown of the approximation. These results indicate that deviations of a few attoseconds (approximately 1-2 as) can be resolved with respect to the predictions of the isotropic asymptotic approximation in the photoelectron energy range of around 13-15~eV.

Similar qualitative considerations also hold true for the experimental data measured with the second-order phase difference $\Delta\varphi_{6,7,9,10}=\Delta\varphi_B$ (Fig.~\ref{Fig3}e,f), with the exception of $\Delta\chi^{(+0,0-)}_{6\omega}$ measured at the highest intensity (Fig.~\ref{Fig3}f). In particular, the error bars for these points are very large when compared to that of other phase differences acquired for the same intensity. We attribute this deviation to possible single-shot fluctuations of the intensity of the harmonic $H_8$ for this phase setting that might affect the fit of the sideband $S^{(0)}_{7,9}$, which overlaps in energy with the photoelectrons released by the absorption of one photon of the harmonic $H_8$. These fluctuations might induce a large variation in the phase of the fit, thus increasing the error bar of the experimental phase differences $\Delta\chi^{(+0,0-)}_{6\omega}$.

Additional measurements were also performed for neon. They generally confirm the breakdown of the isotropic asymptotic approximation and are in qualitative agreement with the measurements in helium. Since the electrons in this case are ionised from the $2p$ shell, the angular momenta involved in the transitions are different from those in helium (see~\ref{Fig4SI} and~\ref{Table8},~\ref{Table9},~\ref{Table10}, and~\ref{Table11}). Consequently, the deviation from the isotropic asymptotic approximation is also modified by the target-dependence of the continuum-continuum phases. The comparison between the experimental data acquired in neon and the corresponding simulations is presented in~\ref{Fig4SI}.

TDSE simulations offer the possibility of disentangling the different contributions to the phase of the emitted photoelectron wave packet and isolating the term that is neglected in the isotropic asymptotic approximation, which is most relevant for the deviations from zero observed in Fig.~\ref{Fig3}. To isolate the phase imparted to the wave packet by the NIR-induced transitions, we consider the phase of the wave packet created in the two-colour photoionisation process, subtracting the Wigner phase obtained in the single-colour photoionisation simulation.
To systematically analyse these deviations, we consider the ideal case presented in Fig.~\ref{Fig1}a, focussing on the continuum-continuum phase between the same energy levels of the continuum (see Fig.~\ref{Fig4}a).
Assuming the validity of the isotropic asymptotic approximation, the sum of the continuum-continuum phases between the two levels corresponds to the phase difference $\Delta\chi_{3\omega}$ of the component oscillating at the frequency $3\omega$ of the adjacent sidebands $S^{(\pm)}_{6,7}$ and $S^{(\pm)}_{9,10}$ (see Fig.~\ref{Fig1}d), as well as to the phase difference $\Delta\chi^{(+0,0-)}_{6\omega}$ of the component oscillating at frequency $6\omega$ between the central sidebands $S^{(0,-)}_{7,9}$ and $S^{(+,0)}_{7,9}$ (see Fig.~\ref{Fig1}e). The main difference with respect to the experimental approach is an additional phase factor of $\pi$ due to the different numbers of photons exchanged in the two pathways that reach the state with the same final energy.

Figure~\ref{Fig4}b presents the continuum-continuum phase $\delta_l$ of the wave function with kinetic energy $E$ and final angular momentum $l=0$ (dashed lines) and $l=2$ (dotted lines) for the hydrogen atom with and without the centrifugal potential term, as obtained for the NIR absorption (red lines) and emission (blue lines) processes. In the first case, we consider the absorption of an XUV photon with energy $E-\hbar\omega$ followed by the absorption of a NIR photon (red solid line). In the second case, we consider the absorption of a photon with energy $E$ and the emission of a NIR photon (blue solid line). This approach allows us to isolate the phase imposed by the interaction with the NIR, which leads to the absorption and the emission of a single NIR photon between the same energy levels of the continuum. We observe that the sum of the two $l$-resolved phases differs from zero (black lines), but approaches this value for increasing kinetic energy of the photoelectron wave packet. Since the influence of the centrifugal potential is neglected in the isotropic asymptotic approximation, the simulations in which this term is removed from the Hamiltonian can be expected to approach the asymptotic results. This is indeed observed, as the sum of the phases extracted from the TDSE without the centrifugal potential (solid black line) is much closer to zero, although it still gives non-zero values due to the additional approximations performed within the isotropic asymptotic approximation (see Ref.~\citenum{dahlstromTheoryAttosecondDelays2013}), and only approaches zero for large photoelectron kinetic energies $E$\@.

Having clearly observed that the centrifugal potential contributes to the deviation from zero, we analyse the role of additional short-range potentials on the deviation. Figure~\ref{Fig4}c shows the sum of the phases obtained by neglecting the centrifugal potential for hydrogen (solid black line) and including the centrifugal term for three different SAE model potentials considering two final angular momentum quantum numbers, $l=0$ (dashed lines) and $l=2$ (dotted lines), corresponding to the absorption and emission of a single NIR photon. We observe that the $l$-resolved evolution of the phase $\delta_l$ does not depend on the specific target. These results suggest that the short-range potential does not affect the deviation from zero of the sum of the continuum-continuum phases.

We analyse the sum of the continuum-continuum phases obtained for transitions characterised by the exchange of two NIR photons using the same approach. This leads to photoelectrons with final angular momenta $l=1$ and $l=3$, as presented in Fig.~\ref{Fig4}d.

The contribution of the centrifugal potential and additional approximations lead to a deviation from zero also for this phase difference, as shown in Fig.~\ref{Fig4}e.
As for the three different potentials considered, i.e., hydrogen, the model potential for He from Ref.~\citenum{tongEmpiricalFormulaStatic2005} (``Tong''), and the effective potential for singly ionised helium ion when the bound electron is frozen in the 1s orbital (``He$^+_{1s}$''), the phase evolutions are superimposed (see Fig.~\ref{Fig4}f), so the influence of the short-range potential is also negligible in this case. Notably, we observe that the variation in absorption and emission does not simply scale linearly with the number of NIR photons exchanged with respect to the case presented in Fig.~\ref{Fig4}a-c. 

\subsection*{Conclusion}
We have analysed the role of the isotropic asymptotic approximation in measuring attosecond time delays in photoionisation in detail. By using a self-referencing approach, we have reported experimental data showing the breakdown of the approximation when comparing the phases of oscillations in different sidebands. The required resolution of a few attoseconds for benchmarking the approximation was achieved because of the large redundancy of information secured by acquiring a complete photoelectron spectrum in a single shot~\cite{marojuAttosecondCoherentControl2023}. Using a solution of the full two electron TDSE and several single-active-electron models, we have shown that the contribution of the centrifugal potential is the dominant term determining the breakdown of the isotropic asymptotic approximation. To a minor extent, the additional approximations inherent in the asymptotic approximation also contribute to the deviation from the experimental values.
As the isotropic asymptotic approximation is fundamental to the interpretation of attosecond time delays in photoionisation, we anticipate that our results will have a significant impact on the community, as they clearly demonstrate the level of accuracy at which the approximation remains valid in attosecond interferometry.

\newpage

\subsubsection*{Acknowledgments}

G.S. acknowledges financial support by FRIAS and by the Deutsche Forschungsgemeinschaft project 547508320 (Project SA 3470/14-1). I.M., M.S, B.M. and G.S. acknowledge financial support by the BMBF project 05K19VF1, the Deutsche Forschungsgemeinschaft project Research Training Group DynCAM (RTG 2717), and Georg H. Endress Foundation. D.B. acknowledges support from the Swedish Research Council grant 2020-06384 and from the Wallenberg Center for Quantum Technology. This work has been financially supported by the European Union's Horizon Europe research and innovation programme under the Marie Skłodowska-Curie grant agreement No 101168628 (project QU-ATTO).
This work has been financially supported by the Swedish Research Council (VR) (grant number 2023-03464) and the Knut and Alice Wallenberg Foundation (grant number 2024.0120) Sweden.
A.P., J.F. and M.B. acknowledge support of the Spanish Ministry of Science, Innovation and Universities-Agencia Estatal de Investigación through Grant Nos.\ PID2021-125894NB-I00, PID2022-138288NB-C32, PID2024-161142NB-I00, CNS2023-145254 and CEX2023001316-M (through the María de Maeztu program for Units of Excellence in R\&D), as well as the Comunidad de Madrid PRICIT program through project PCD-IP3PCD026. We thank Morgan Berkane and Christoph Dittel for careful reading of the manuscript and for stimulating discussions.

\subsubsection*{Author Contributions Statement}
P.K.M. and S.R. analysed the data. M.B.L, J.F., A.P. and G.S contributed to the analysis of the phases of the sideband oscillations.
P.K.M., M.D.F., O.P., D.B., I.M., M.S., T.C., K.P., J.M.,C.C. and G.S. participated to the beamtime and acquired the experimental data. L.G., E.A., P.R.R., and G.D.N. optimised the operation of the FEL during the beamtime. M.Z. was responsible for the alignment of the PADRES system and for the focusing of the XUV harmonics. A.D and M.D. were responsible for the alignment of NIR laser and for its optimisation. R.J.S. and R.F. were responsible for the magnetic bottle photoelectron spectrometer. L.P. and F.F. constructed and optimised the XUV spectrometer. M.B.L, J.F. and A.P. performed the TDSE and SAE simulations. K.S. contributed to the initial simulations of the project. P.K.M., M.B.L, J.F., A.P., and G.S. interpreted the experimental data. P.K.M. and G.S. conceived the idea of the experiment. G.S. supervised the work and wrote the manuscript, which was discussed and agreed on by all coauthors.\\

\subsubsection*{Competing Interests Statement}
The authors declare no competing interests.\\
\subsubsection*{Correspondence and requests for materials}
Correspondence and requests for materials should be addressed to johannes.feist@uam.es, alicia.palacios@uam.es, and giuseppe.sansone@physik.uni-freiburg.de\\

\section*{Methods}

\subsection*{Methods}
\subsubsection*{Experimental Methods}\label{Methods_EM}

An ultraviolet laser pulse ($\lambda=266$ nm) was used in combination with an undulator to create a periodic density modulation (bunching) at the laser wavelength in the electron bunch of the FERMI free electron laser operating at 50 Hz repetition rate. The six downstream undulators constituting the FERMI FEL-1 radiator were tuned to different harmonics of the seed laser wavelength: U1-U2 at harmonic $H_{10}$, U3-U4 at harmonic $H_9$, U5 at $H_7$, and U6 at $H_6$. Relative phase between different harmonics was controlled using dedicated phase shifter available between undulators. The typical pulse energy of each harmonic emitted by resonant undulators was on the order of a few microjoules ($\mu$J). However, the  bunched electron beam can potentially emit coherently at all harmonics in any magnetic element of the electron beamline downstream the bunching section, thus leading to a spurious and small signal at other harmonics. In the experiment a small signal at $H_8$ was measured, which, however, was estimated to be two to three orders of magnitude weaker than the emission produced in the resonant undulators. 
The XUV harmonics were transported towards the low-density matter end station~\cite{lyamayevModularEndstationAtomic2013} using the PADReS system~\cite{zangrandoRecentResultsPADReS2015}. They were focused into the experimental end station in a helium gas target using the KAOS active optics system~\cite{raimondiKirkpatrickBaezActiveOptics2019}.
The XUV harmonics were collinearly recombined with the NIR field using a drilled mirror. The photoelectrons emitted in the two-colour photoionisation process were collected in a solid angle of $2\pi$ srad along the common polarisation direction of the fields by a magnetic bottle electron spectrometer. 
Photoelectron spectra were measured on a single-shot basis. The relative phase between the two groups of harmonics ($H_{10}$-$H_9$) and ($H_7$-$H_6$) was tuned to the value of $\approx\pm\pi/2$ to ensure the lowest error using the attosecond timing tool. The second-order phase was adjusted by operating on the phase shifter placed between the undulators U5 and U6 thus affecting the phase of the harmonic $H_6$. After the interaction region, the XUV radiation was analysed using an XUV spectrometer that records the intensity of each harmonic on a single-shot basis~\cite{zeniCompactSpectrometerDedicated2025}.
For each experimental configuration (corresponding to a fixed relative phase of the harmonics and intensity of the NIR pulses), we recorded $10000$ single-shot photoelectron spectra; periodic gas shots were excluded for background subtraction, yielding $ 8571$ signal shots used in the analysis.

\subsubsection*{Data Analysis Methods}\label{Methods_DAM}

The attosecond timing tool was used to reorder each photoelectron spectrum acquired with the magnetic bottle spectrometer according to the relative phase between the attosecond pulse train and the NIR field. The single-shot photoelectron spectra used in the inversion were selected by applying conditions on the intensities of the harmonics measured with the XUV spectrometer for the corresponding shot. In particular, we selected only spectra for which the intensity of each individual harmonic lies between $\pm10\%$ of the mean value for that particular condition. The single-shot photoelectron spectra were reordered according to the phase $\Phi=3\omega\tau+\eta$, where $\tau$ indicates the relative delay between the XUV and NIR fields and $\eta=(\varphi_{10}+\varphi_9)-(\varphi_7+\varphi_6)$~\cite{marojuAttosecondPulseShaping2020, marojuAnalysisTwocolorPhotoelectron2021, marojuComplexAttosecondWaveform2021}.
The sideband oscillations were fitted using a double cosine function according to Eq.~\ref{Eq1}. The error in the fitting procedure is the error in the estimation of the phases $\chi_{3\omega}^{\mathrm{exp}}$ and $\chi_{6\omega}^{\mathrm{exp}}$ for each sideband.

\clearpage

\begin{figure}[!htbp]
\centering \resizebox{1.0\hsize}{!}{\includegraphics{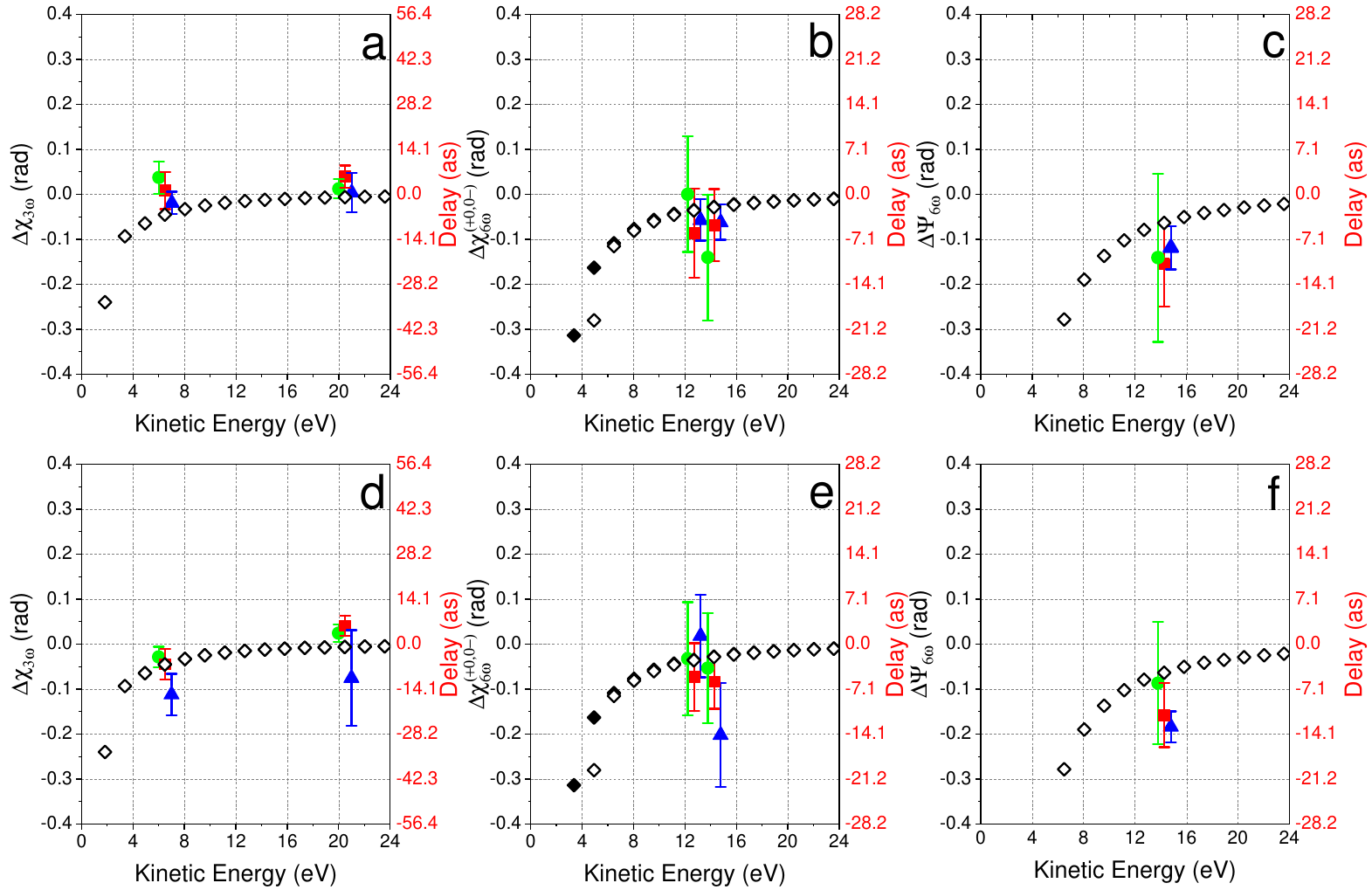}}
\caption{\textbf{Comparison between simulations and experimental results for the phase differences $\Delta\chi_{3\omega}$, $\Delta\chi^{(+0,0-)}_{6\omega}$, and $\Delta\Psi_{6\omega}$ in helium.} Comparison of the theoretical simulations (diamonds) obtained using the TDSE and the experimental data (coloured symbols) for the phase differences $\Delta\chi_{3\omega}$ (a,d), $\Delta\chi^{(+0,0-)}_{6\omega}$ (b,e), and $\Delta\Psi_{6\omega}$ (c,f) for three different NIR intensities: $I_1=1.2\times10^{12}$~W/cm$^2$ (green circles), $I_2=2.2\times10^{12}$~W/cm$^2$ (red squares), and $I_3=4.0\times10^{12}$~W/cm$^2$ (blue triangles). The second-order phase difference $\Delta\varphi_{6,7,9,10}$ for the measurements presented in panels (a-c) and (d-f) were $\Delta\varphi_A$ and $\Delta\varphi_B$, respectively. The experimental data were corrected for the contribution of the harmonic $H_8$. The error bars are derived by propagation of the uncertainties in the fitting procedure of the sideband oscillations. The open (full) diamonds in panels (b,e) indicate the phase difference $\Delta\chi_{6\omega}^{(+0)}$ ($\Delta\chi_{6\omega}^{(0-)}$).}
\label{Fig3}
\end{figure}

\begin{figure}[!htbp]
\centering \resizebox{1.0\hsize}{!}{\includegraphics{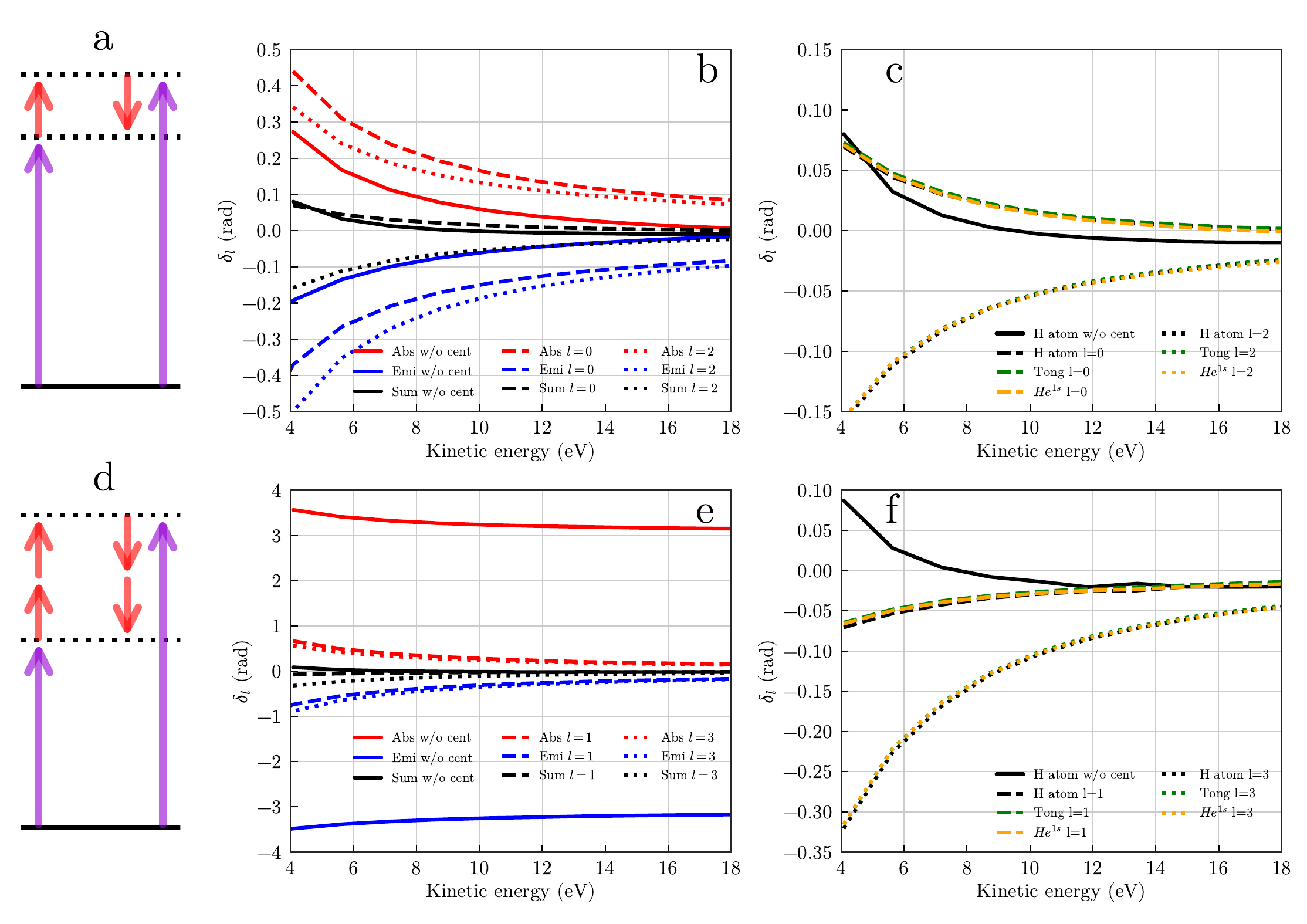}}
\caption{\textbf{Origin of the breakdown of the isotropic asymptotic approximation.} a) Scheme employed in b) and c) to obtain the continuum-continuum phases. The Wigner phase has already been subtracted in b) and c). b) Continuum-continuum phase $\delta_l$ obtained by solution of the TDSE for the hydrogen atom neglecting the centrifugal potential (solid lines) and with the centrifugal potential for final angular momentum $l=0$ (dashed lines) and $l=2$ (dotted lines), for absorption (red lines), emission (blue lines), and the resulting sum of absorption and emission lines (black lines) in the two-colour photoionisation process of hydrogen. c) Sum of absorption and emission phases for $l=0$ (dashed lines) and $l=2$ (dotted lines) for the hydrogen atom (black lines) and two helium SAE model potentials: ``Tong'' (green lines) and ``He$^+_{1s}$'' (orange), see main text. The result for hydrogen atom without the centrifugal potential (solid black line) is also shown.  d) Scheme employed in e) and f) where the phases correspond to those of two NIR photon transitions, the Wigner phase has been subtracted as for b) and c). e) Two-photon continuum-continuum phase without centrifugal potential (solid lines) and for $l=1$ (dashed lines) and $l=3$ (dotted lines) for the same potentials as c. f) same as c), but for the two-photon transitions. The intensity of the XUV and IR radiation used in the simulation was $10^{11}$~W/cm$^2$ and $10^{10}$~W/cm$^2$, respectively. The pulses have a $\sin^8$ envelope with a total duration of 40 fs.
}
\label{Fig4}
\end{figure}

%\begin{thebibliography}{10}
%\bibliographystyle{unsrt}
\bibliographystyle{naturemag}
%\printbibliography
%\bibliographystyle{unsrt}
\bibliography{library.bib}

\clearpage
\newpage
\setcounter{figure}{0}
\setcounter{table}{0}
%\DeclareCaptionLabelFormat{figure}{\makebox[2.5cm][l]{Fig. #2:}}
\renewcommand{\thefigure}{Extended Data Fig. \arabic{figure}}
\renewcommand{\thetable}{Extended Data Table \arabic{table}}
\renewcommand{\theequation}{Eq. SI\arabic{equation}}

\captionsetup[figure]{name={}}
\captionsetup[table]{name={}}

\setcounter{page}{1}
\setcounter{section}{1}
\clearpage
\preto{\section}{\resetlinenumber}

\setcounter{page}{1}
\setcounter{section}{1}
\setcounter{equation}{0}

\section*{Supplementary Information}

\subsection*{Experimental information on XUV spectrum characterisation}
In the experiment, the phase difference between the pairs of harmonics $H_6$-$H_7$ and $H_9$-$H_{10}$ was adjusted to meet the condition
\begin{equation}
 \Delta\varphi_{6,7,9,10}=(\varphi_{10}-\varphi_9)-(\varphi_7-\varphi_6)\approx \mp\pi/2=\Delta\varphi_{A,B}
\end{equation}
for which the highest accuracy in the reconstruction of the sideband oscillations is achieved~\cite{marojuAttosecondCoherentControl2023}.
The second-order phase differences of the different harmonics (proportional to the group delay dispersion) were determined either by using the correlation method or by using the attosecond timing tool approach. Both methods return consistent results~\cite{marojuAttosecondTemporalStructure2025}.
The phase of the harmonic $H_8$ was estimated by considering its effect on the phase of the oscillations of the central sidebands $S^{(-,+)}_{7,9}$ and optimising the matching with the experimental data by a fitting procedure. The procedure will be described in detail in the next section.

The amplitudes ($F_q$) of the harmonics were estimated using two different procedures. In the first one, the intensities of the harmonics were acquired on a single-shot basis using an XUV spectrometer placed after the interaction region. The intensities of the harmonics were corrected for the reflectivity of a gold mirror used to direct the radiation to the entrance slit of the spectrometer (see~\ref{Table1}). In the second one, the amplitudes of the single harmonics were obtained using the same fitting procedure described in the next section.  

\begin{table}[h]
  \centering
  \begin{tabular}{| c | c | c | c | c | c | c |}
    \hline
      & $H_{10}$ & $H_9$ & $H_8$ &$H_7$ & $H_6$ \\ \hline
     $\hbar\omega_q$ (eV) & $46.96\pm0.05$ & $42.27\pm0.04$ & $37.57\pm0.04$ & $32.88\pm0.03$ & $28.18\pm0.03$ \\\hline
      \multicolumn{6
      } {|c|} {\textbf{$\Delta\varphi_{6,7,9,10}=\Delta\varphi_A=4.722\pm0.009$} (rad)} \\ \hline
    $F_q$ (arb. un.)      &      $0.40\pm0.03$      &  $0.80\pm0.03$              & $0.11\pm0.01$ & $1.0\pm0.14$               & $0.75\pm0.06$  \\\hline\hline
          \multicolumn{6
      } {|c|} {\textbf{$\Delta\varphi_{6,7,9,10}=\Delta\varphi_B=1.418\pm0.009$} (rad)} \\ \hline
     $F_q$ (arb. un.)      &       $0.40\pm0.03$      & $0.78\pm0.03$               & $0.10\pm0.01$ &  $1.0\pm0.13$               & $0.72\pm0.06$  \\\hline
\end{tabular}
  \caption{Photon energies ($\hbar\omega_{q}$) and amplitudes ($F_q$) of the five harmonics for the two harmonic configuration characterised by $\Delta\varphi_{6,7,9,10}=\Delta\varphi_{A}$ and $\Delta\varphi_{6,7,9,10}=\Delta\varphi_{B}$, respectively.
  The error bars on the field amplitudes were estimated considering the error propagation of the standard deviation of the single-shot harmonic intensities.}\label{Table1}
\end{table}

\subsection*{Determination of the value of the amplitude and phase of the harmonics using a fitting procedure}
The amplitudes and phases of all harmonics, including those of the harmonic $H_8$, were determined by a global fitting of all sideband oscillations using the results of the TDSE simulations. The results of the global fitting return values in close agreement with those extracted from the determination of second-order phase difference based on the strong-field-approximation (SFA) and the independent measurement of the harmonic amplitudes.
The results are presented in~\ref{Table2}.

\begin{table}[h]
  \centering
  \begin{tabular}{| c | c | c | c | c | c | c |}
    \hline
      & $H_{10}$ & $H_9$ & $H_8$ &$H_7$ & $H_6$ \\ \hline
      \multicolumn{6
      } {|c|} {$\Delta\varphi_{6,7,9,10}=\Delta\varphi_A=5.02\pm0.26$~rad }\\ \hline
    $F_q$ (arb. un.)       &      $1.26\pm0.36$      &  $1.21\pm0.21$              & $0.11\pm0.06$ & $1.00\pm0.16$               & $0.78\pm0.14$  \\\hline
    $\varphi_q$ (arb. un.) &      $4.00\pm0.11$      &  $3.38\pm0.10$              & $4.91\pm0.16$ & $1.89\pm0.05$               & $0.00\pm0.00$  \\\hline\hline
          \multicolumn{6
      } {|c|} { $\Delta\varphi_{6,7,9,10}=\Delta\varphi_B=1.91\pm0.30$~rad }\\ \hline
     $F_q$ (arb. un.)      &       $0.99\pm0.29$      & $1.11\pm0.17$               & $0.09\pm0.05$ &  $1.0\pm0.16$               & $0.76\pm0.14$  \\\hline
     $\varphi_q$ (arb. un.) &      $0.68\pm0.13$      &  $4.75\pm0.12$              & $4.80\pm0.16$ & $0.39\pm0.05$               & $0.00\pm0.00$  \\\hline\hline
\end{tabular}
  \caption{Amplitudes ($F_q$) and phases ($\varphi_q$) of the five harmonics for the two harmonic configuration characterised by $\Delta\varphi_{6,7,9,10}=\Delta\varphi_{A}$ (upper three lines) and $\Delta\varphi_{6,7,9,10}=\Delta\varphi_{B}$ (lower three lines) obtained using the TDSE-based global fitting procedure (see text), respectively. The fit also returned the values of the phases $\Delta\varphi_{A}$ and $\Delta\varphi_{B}$.
  The error bars on the field amplitudes and phases were derived from the fitting procedure.}\label{Table2}
\end{table}
The values of the amplitudes of the harmonics $H_6$, $H_7$, and $H_8$ reported in~\ref{Table1} are in excellent agreement with those extracted from the fitting procedure shown in~\ref{Table2}. The agreement is less good for the amplitudes of the harmonics $H_9$ and $H_{10}$. We attribute this discrepancy to the energy resolution of the magnetic bottle electron spectrometer, which leads to a partial overlap of the photoelectron peaks of the sidebands and harmonics in the high photoelectron kinetic energy region. The limited resolution results in an increase in the sideband signal estimated from the experimental data, due to the partial contribution of the harmonic signal, thus leading to an overestimation of the harmonic intensities in the fitting procedure.

The values of the second-order phase differences are in good agreement with those extracted from the independent approach based on the SFA analysis of the sideband oscillations~\cite{marojuAttosecondTemporalStructure2025}, as shown in~\ref{Table3}.
\begin{table}[h]
  \centering
%\framebox[1.1\width]{
  \begin{tabular}{| c | c | c | c | c | c | c | p{8cm} |}
    \hline
   Configuration & Model & $\Delta\varphi_{6,7,9,10}$ (rad) & $\Delta\varphi_{8,9,10}$ (rad) &  $\Delta\varphi_{7,8,9}$ (rad) &  $\Delta\varphi_{6,7,8}$ (rad) \\\hline
     \multirow{2}{*} { $\Delta\varphi_A$} & SFA  & $4.722\pm0.009$               & $1.96\pm0.15$      &      $1.78\pm0.15$      &  $1.00\pm0.10$  \\
    & TDSE & $5.02\pm0.26$ & $2.16\pm0.44$      &      $1.72\pm0.09$      &  $1.14\pm0.05$  \\\hline\hline\hline
   \multirow{2}{*} { $\Delta\varphi_B$} & SFA  & $1.418\pm0.009$               & $2.07\pm0.15$      &      $1.61\pm0.15$      &  $4.10\pm0.10$  \\
    & TDSE & $1.91\pm0.30$  & $2.25\pm0.05$      &      $1.73\pm0.09$      &  $4.21\pm0.05$  \\\hline
\end{tabular}
%} \par
  \caption{Second-order phase differences for the two configurations corresponding to different values of $\Delta\varphi_{6,7,9,10}$ obtained using the SFA model and the TDSE fitting procedure}\label{Table3}
\end{table}
The values obtained following the SFA-based approach and the TDSE fitting procedure present an offset of about 0.1-0.2~rad (up to 0.5~rad for $\Delta\varphi_{6,7,9,10}$) that we attribute to the atomic photoionisation phase, whose contribution is not accounted for in the SFA model.

\subsection*{Wave function synthesis method}
We use a wave function synthesis method to obtain the sideband oscillation phases in the RABBIT spectrograms for arbitrary values of the NIR and XUV phases and intensities without needing to solve the TDSE for each set of parameters. Instead, for each NIR intensity of interest, we perform only a single TDSE simulation for each harmonic in the presence of the NIR field, and then we reconstruct the final wave function for arbitrary values of the NIR phase and XUV phases and intensities by synthesising the wave function from the individual harmonic + NIR TDSE simulations. This procedure relies on the perturbative nature of the XUV-NIR interaction in the regime considered here, where the XUV intensities are low enough to ensure that only single-photon processes contribute in the energy region of interest, and on the rotating-wave approximation that ensures that photon absorption and stimulated emission processes have a well-defined dependence on the field phases. For concreteness, we write the electric field as
\begin{equation}
  E(t) = \sum_a F_a f_a(t) e^{-i\omega_a t - i\varphi_a} + c.c.
\end{equation}
where $E(t)$ is a scalar quantity since we assume linear polarisation along a fixed direction; $a$ labels the fields (NIR and harmonics), $f_a(t)$ is the pulse envelope (with a maximum value $|f_a(t)|=1$), $\omega_a$ is the carrier frequency, and $F_a$ and $\varphi_a$ are the field amplitude and phase.

For the purposes of wave function synthesis, we actually perform a scan of the \emph{phase} of the NIR field ($\varphi_{\mathrm{NIR}}$) rather than the delay between the XUV and NIR fields. These two alternatives are equivalent for sufficiently long pulses, but the phase scan has several advantages in the theoretical description. First, since $\varphi_\mathrm{NIR}$ and $\varphi_\mathrm{NIR}+2\pi$ describe the same field, this trivially makes all observables (such as the sideband intensities) periodic functions of $\varphi_\mathrm{NIR}$, such that, by Fourier analysis, all observables can only oscillate at integer multiples of this phase, $O(\varphi_\mathrm{NIR}) = \sum_n O_n e^{-i n \varphi_\mathrm{NIR}}$, which corresponds to oscillation at integer multiples of the NIR frequency $\omega_\mathrm{NIR}$ when expressed through the equivalent time delay. Second, the phase dependence of the different multiphoton pathways can be easily tracked analytically, as described below, allowing the synthesis of the full phase scan from a single simulation (or, equivalently, direct calculation of the oscillation amplitudes and phases without requiring an explicit phase scan).

The synthesis procedure begins by expanding the final wave function after the interaction with the XUV and NIR fields into a perturbation series in the XUV (harmonic) field amplitudes $F_{H_m}$:
\begin{equation}
    \ket{\Psi(t_f)} = \ket{\Psi_0(t_f)} + \sum_m F_{H_m} e^{-i \varphi_{H_m}} \ket{\Psi_{H_m}(t_f)} + \sum_m F_{H_m} e^{i \varphi_{H_m}} \ket{\Psi'_{H_m}(t_f)} + O(F_{H_m}^2),
    \label{WF_harmonics_contribution}
\end{equation}
where $\ket{\Psi_0(t_F)}$ is the wave function under the action of the NIR field only, which has negligible ionisation and thus does not contribute to our observable of interest. We restrict the expansion to first order in $F_{H_m}$ since we are only interested in one-photon single ionisation. 
The terms $\ket{\Psi_{H_m}(t_F)}$ and $\ket{\Psi'_{H_m}(t_F)}$ correspond to the absorption and stimulated emission of a photon from harmonic $H_m$, respectively (i.e., transitions induced by the terms $\propto e^{-i\omega_{H_m}t}$ and $\propto e^{i\omega_{H_m}t}$ in $E(t)$). Stimulated emission processes lead to negligible ionisation probability since the initial state is the ground state; thus, so we can neglect the terms $\propto\ket{\Psi'_{H_m}(t_F)}$ in the following. Each term $\ket{\Psi_{H_m}(t_F)}$ can then be obtained by solving the TDSE with only harmonic $H_m$ and the NIR field present. Changing the intensities or phases of the harmonics $F_{H_m}$ then only requires multiplying the corresponding term in the expansion by the appropriate factor. We stress that the perturbation expansion is only performed in the harmonic XUV fields, and the results contain all orders of interaction with the NIR field (i.e., absorption and emission of arbitrary numbers of NIR photons).

\begin{figure}[tbp]
    \centering
    \includegraphics[width=\linewidth]{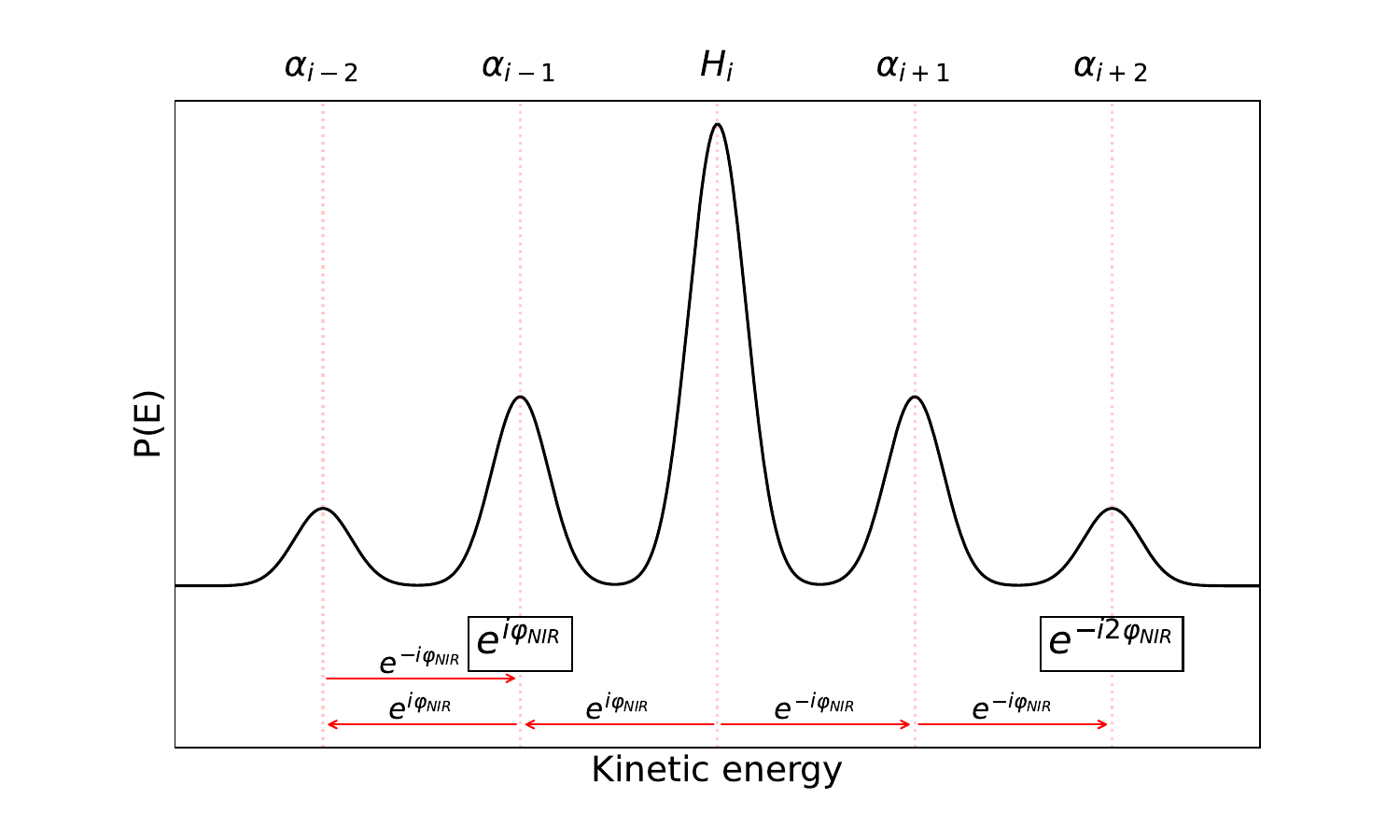}
    \caption{\textbf{Schematic representation of the sideband spectrum created by absorption or emission of NIR photons from the mainband created by harmonic $H_i$}. The phases corresponding to each NIR photon emission and absorption event are indicated below, showing that sideband $\alpha_{i+n}$ obtains complex phase $e^{-i n \varphi_\mathrm{NIR}}$.}
    \label{Fig0SI}
\end{figure}

We now focus on the single ionisation subspace of interest here and show that the dependence on the NIR phase $\varphi_\mathrm{NIR}$ can also be synthesised by multiplying the single ionisation amplitudes by a phase-dependent factor. For each simulation with a single harmonic $H_m$ and the NIR field, the final wave function contains contributions from the main photoelectron band created by the absorption of one XUV photon from harmonic $H_m$, as well as sidebands created by the absorption or emission of NIR photons from this main band, as depicted in \ref{Fig0SI}.
Since we are using relatively intense NIR fields, the interaction with the NIR field is not necessarily described by lowest-order perturbation theory; for example, the first sideband next to each harmonic contains not only contributions from single-NIR-photon transitions but also from three-NIR-photon ones (e.g., two absorptions and one emission). However, as long as the sidebands are still well-separated in energy, the ``net'' NIR photon number (difference of absorbed and emitted photons) to reach each sideband from the main harmonic band is well-defined. Within the rotating-wave approximation (i.e., neglecting counter-rotating terms that do not conserve energy), each absorption of a NIR photon contributes a phase factor $e^{-i\varphi_\mathrm{NIR}}$, while each emission contributes a factor $e^{i\varphi_\mathrm{NIR}}$, such that the total phase contribution from the NIR field to each sideband is determined solely by the net number of NIR photons absorbed or emitted to reach that sideband from the main harmonic band (see \ref{Fig0SI}). 
Hence, within the energy range of each sideband, we can simply multiply the ionisation amplitudes by the corresponding phase factor to obtain the correct dependence on $\varphi_\mathrm{NIR}$. 

In order to obtain the sideband oscillation amplitudes and phases from the synthesis, we thus split the wave function into components within each sideband $N$, i.e., $\ket{\Psi_{H_m}(t_f)} = \sum_{N=0}^\infty \ket{\Psi_{H_m}^{(N)}(t_f)}$, where $\ket{\Psi_{H_m}^{(N)}(t_f)}$ contains all final states with an energy of $E_N = E_0 + (N \pm \frac12) \hbar\omega_\mathrm{NIR}$, where $E_0$ is the ground-state energy, such that $N$ labels the central energy of the sideband. Due to the discussion above, the net number of NIR photons absorbed or emitted to reach each sideband from the main band generated by harmonic $H_m$ is well-defined and given by $n_m = N - 3m$.
The total wave function within sideband $N$ for arbitrary harmonic amplitudes and NIR phases can then be written as
\begin{equation}
    \ket{\Psi^{(N)}(t_f)} = \sum_m F_{H_m} e^{-i \varphi_{H_m}} e^{-i n_m \varphi_\mathrm{NIR}} \ket{\Psi_{H_m}^{(N)}(t_f)}.
    \label{eq:sideband_synthesis}
\end{equation}
Introducing the projection operator $\hat{P}_{\mathrm{U}}$ for single ionisation into the upper half-space, the corresponding sideband amplitudes are then given by:
\begin{equation}
  P_{\mathrm{U}}^{(N)} = \braket{\Psi^{(N)}(t_f) | \hat{P}_{\mathrm{U}} | \Psi^{(N)}(t_f)} = \sum_{m,m'} F_{H_m} F_{H_{m'}} e^{i(\varphi_{H_{m'}} - \varphi_{H_m})} e^{i(n_{m'} - n_{m}) \varphi_\mathrm{NIR}} C_{m,m'}^{(N)},
\end{equation}
where $C_{m,m'}^{(N)} = \braket{\Psi_{H_{m'}}^{(N)}(t_f) | \hat{P}_{\mathrm{U}} | \Psi_{H_{m}}^{(N)}(t_f)}$ are complex amplitudes that can be obtained directly from the TDSE simulations with individual harmonics and the NIR field (with phase $\varphi_\mathrm{NIR} = 0$). Taking into account that the diagonal parts $m=m'$ are real, that exchanging $m$ and $m'$ corresponds to complex conjugation, and that $n_{m'}-n_{m} = 3(m-m')$, the sideband amplitudes can be expressed as
\begin{equation}
  \begin{split}
  P_{\mathrm{U}}^{(N)} &= \sum_{m} F_{H_m}^2 C_{m,m}^{(N)} + 2\Re\left[ \sum_{m>m'} F_{H_m} F_{H_{m'}} e^{i(\varphi_{H_{m'}} - \varphi_{H_m})} e^{3i(m - m') \varphi_\mathrm{NIR}} C_{m,m'}^{(N)} \right] =\\
  &= c_0^{(N)} + 2 \Re\left[e^{3i \varphi_\mathrm{NIR}} c_{3}^{(N)}\right] + 2 \Re\left[e^{6i \varphi_\mathrm{NIR}}c_{6}^{(N)}\right] + \ldots,\\
  &= c_0^{(N)} + 2 |c_{3}^{(N)}| \cos(3\varphi_\mathrm{NIR} + \arg{c_{3}^{(N)}}) + 2 |c_{6}^{(N)}| \cos(6\varphi_\mathrm{NIR} + \arg{c_{6}^{(N)}}) + \ldots,
  \end{split}
\end{equation}
where the complex coefficients $c_{k}^{(N)}$ contain all contributions to sideband $N$ oscillating at frequency $k \omega_\mathrm{NIR}$, i.e., with $m-m' = k$. The sideband oscillation amplitudes and phases can be extracted directly from this expression without requiring an explicit phase scan and converted to oscillations in terms of the equivalent delay $\tau$ by using $\varphi_\mathrm{NIR} = \omega_\mathrm{NIR} \tau$. This analysis shows that the interference between a pair of harmonics $H_m$ and $H_{m'}$ always contributes a term oscillating with frequency $3(m-m')\omega_\mathrm{NIR}$, i.e., interference between adjacent harmonics ($m = m'+1$) contributes oscillations at frequency $3\omega_\mathrm{NIR}$, interference between harmonics separated by one harmonic order ($m = m'+2$) contributes oscillations at frequency $6\omega_\mathrm{NIR}$, and so on.

Compared to explicit delay scans that include all harmonics and the NIR field, the synthesis method has two major advantages in the context of the present work: First, it allows us to efficiently explore a large parameter space of harmonic amplitudes and phases without needing to solve the TDSE for each set of parameters, which would be computationally prohibitive. This efficiency is crucial for fitting the experimental results by adjusting the harmonic amplitudes and phases. Second, it allows us to isolate the contributions of specific pairs of harmonics to each sideband by restricting the sums in~\ref{eq:sideband_synthesis} to only those harmonics of interest, which is not possible in an explicit TDSE simulation with all harmonics present. This is particularly useful here to isolate the contribution of the spurious harmonic 8 to the various phases of interest, as discussed in the main text.

\subsection*{Effect of the harmonic $H_8$ on the phase differences between adjacent and non-adjacent sidebands}
\begin{figure}[ht]
\centering \resizebox{1.0\hsize}{!}{\includegraphics{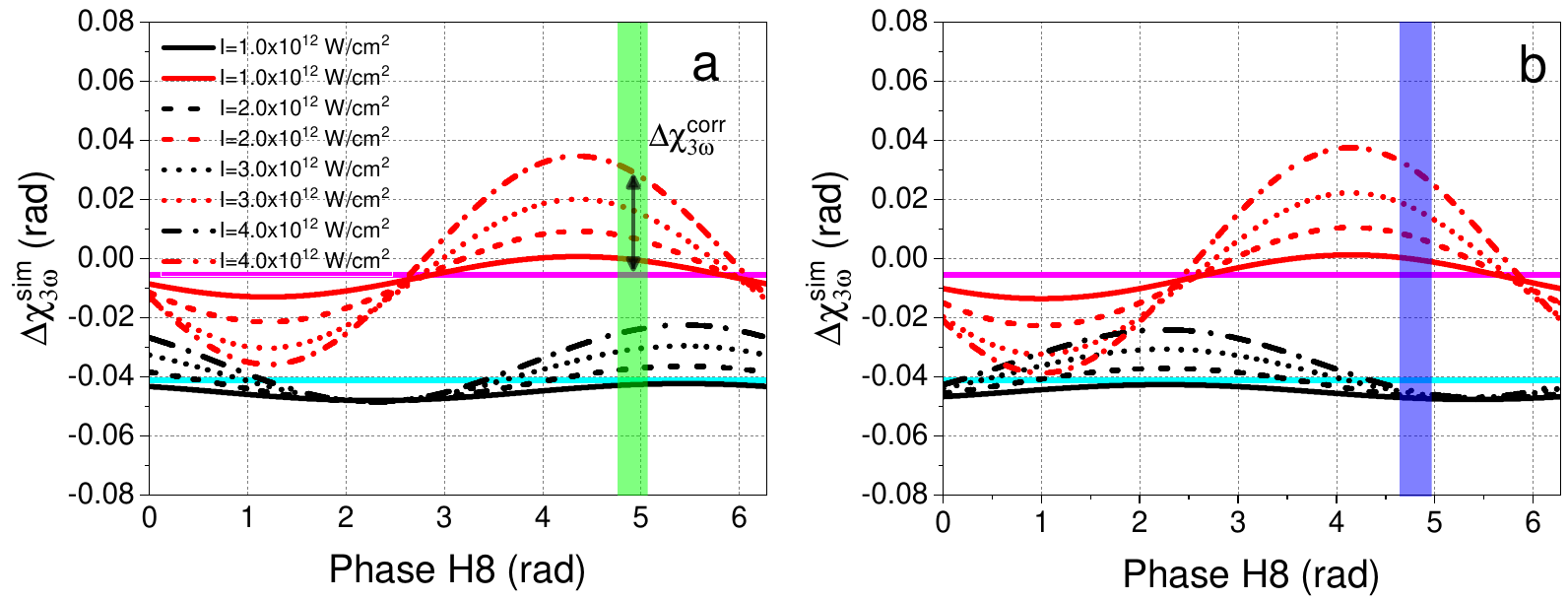}}
\centering \resizebox{1.0\hsize}{!}{\includegraphics{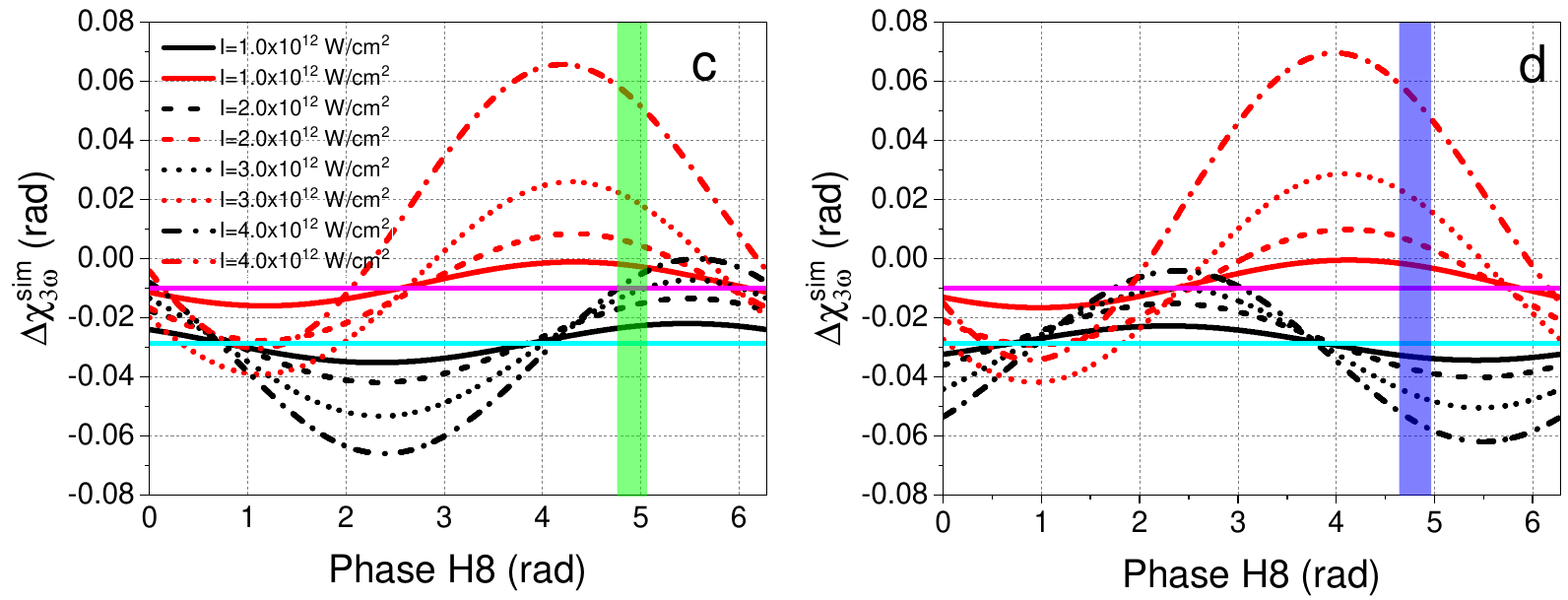}}
\caption{\textbf{Effect of the harmonic $H_8$ on the simulated phase difference $\Delta\chi^{\mathrm{sim}}_{3\omega}$.} Evolution of the phase $\Delta\chi^{\mathrm{sim}}_{3\omega}$ for different energies of the emitted photoelectron (E$_1$ (red curves, panels a,b); E$_2$ (black curves, panels a,b); (E$_a$ (red curves, panels c,d); E$_b$ (black curves, panels c,d)), for the four different intensities  $I=1.0\times10^{12}$~W/cm$^2$, $2.0\times10^{12}$~W/cm$^2$, $3.0\times10^{12}$~W/cm$^2$, and $4.0\times10^{12}$~W/cm$^2$, and for the two second-order phase difference $\Delta\varphi_{6,7,9,10}=\Delta\varphi_A$ (a,c) and $\Delta\varphi_B$ (b,d). The green (blue) shaded area indicate the interval corresponding to the best estimation of the value of the phase of the harmonic $H_8$. The cyan (magenta) line indicates the value of $\Delta\chi^{\mathrm{sim}}_{3\omega}$ obtained without the contribution of the harmonic $H_8$ ($F_8=0$) for the energy $E_1$ ($E_2$) (a,b) and $E_a$ ($E_b$) (c,d). In panel (a), the term $\Delta\chi^{\mathrm{corr}}_{3\omega}$ defined in~\ref{corr} is indicated. Panels a and b: simulations in helium. Panels c and d: simulations in neon.}
\label{Fig1SI}
\end{figure}

\begin{figure}[ht]
\centering \resizebox{1.0\hsize}{!}{\includegraphics{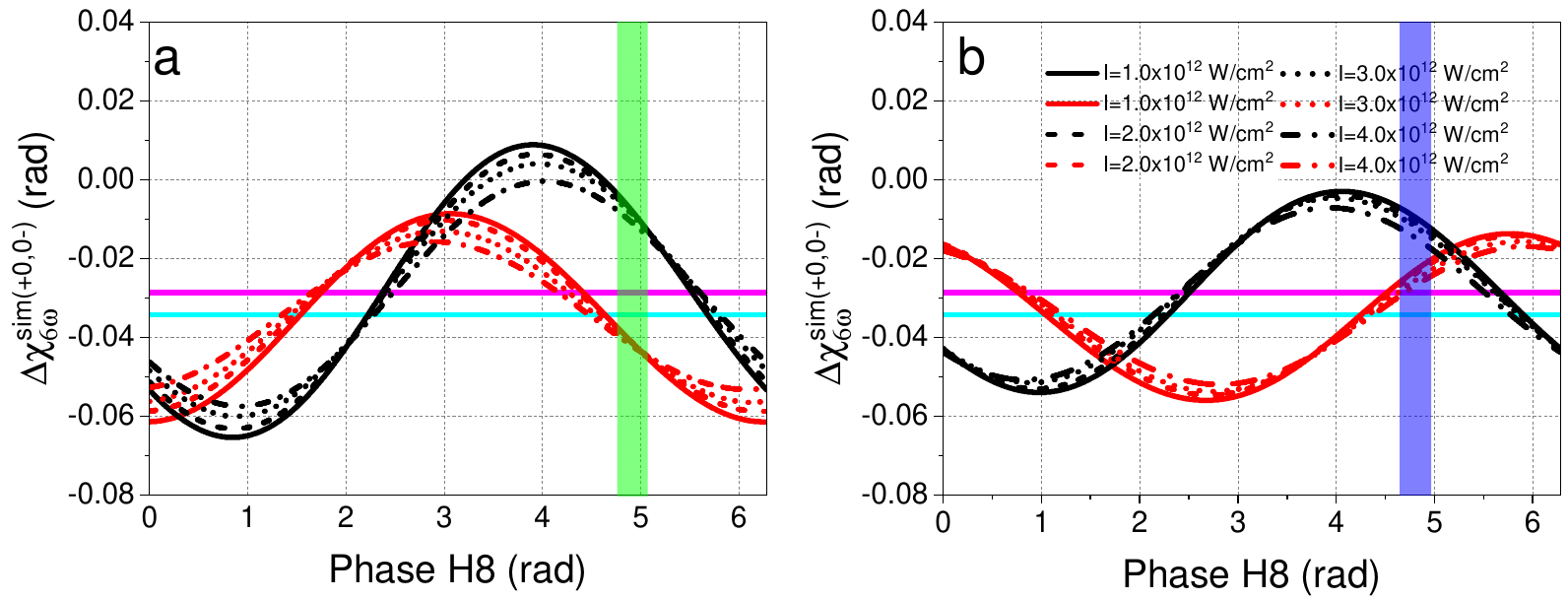}}
\centering \resizebox{1.0\hsize}{!}{\includegraphics{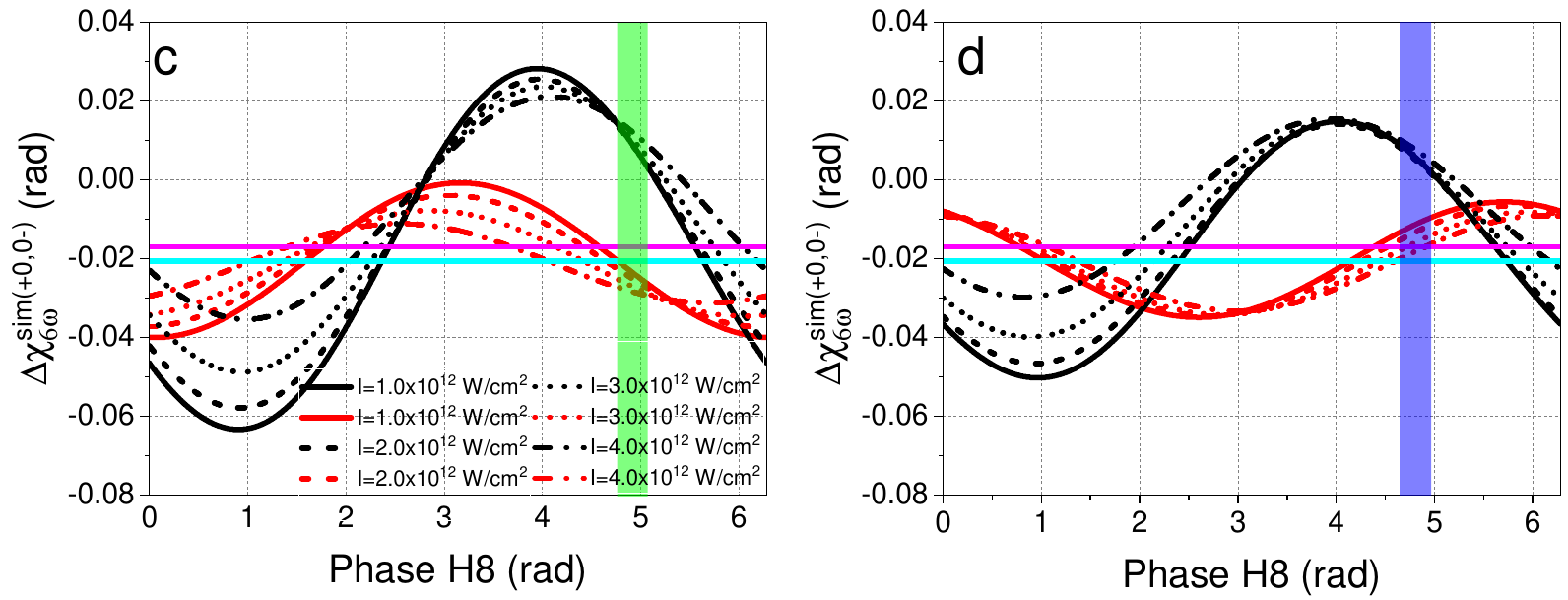}}
\caption{\textbf{Effect of the harmonic $H_8$ on the simulated phase differences $\Delta\chi^{\mathrm{sim};(+0,0-)}_{6\omega}$.} Evolution of the phase $\Delta\chi^{\mathrm{sim};(0-)}_{6\omega}$ (red lines) and $\Delta\chi^{\mathrm{sim};(+0)}_{6\omega}$ (black lines) for different energies of the emitted photoelectron ($\overline{E}_1$ (red curves, panels a,b) and $\overline{E}^*_1$ (black curves, panels a,b); $\overline{E}_a$ (red curves, panels c,d) and $\overline{E}^*_a$ (black curves, panels c,d)), for the four different intensities  $I=1.0\times10^{12}$~W/cm$^2$, $2.0\times10^{12}$~W/cm$^2$, $3.0\times10^{12}$~W/cm$^2$, and $4.0\times10^{12}$~W/cm$^2$, and for the two second-order phase difference $\Delta\varphi_{6,7,9,10}=\Delta\varphi_A$ (a,c) and $\Delta\varphi_B$ (b,d). The green (blue) shaded area indicate the interval corresponding to the best estimation of the value of the phase of the harmonic $H_8$. The cyan (magenta) line indicates the value of $\Delta\chi^{\mathrm{sim};(+0,0-)}_{6\omega}$ obtained without the contribution of the harmonic $H_8$ ($F_8=0$) for the energy $\overline{E}_1$ ($\overline{E}^*_1$) (a,b) and  $\overline{E}_a$ ($\overline{E}^*_a$) (c,d). Panels a and b: simulations in helium. Panels c and d: simulations in neon.}
\label{Fig2SI}
\end{figure}

\begin{figure}[ht]
\centering \resizebox{1.0\hsize}{!}{\includegraphics{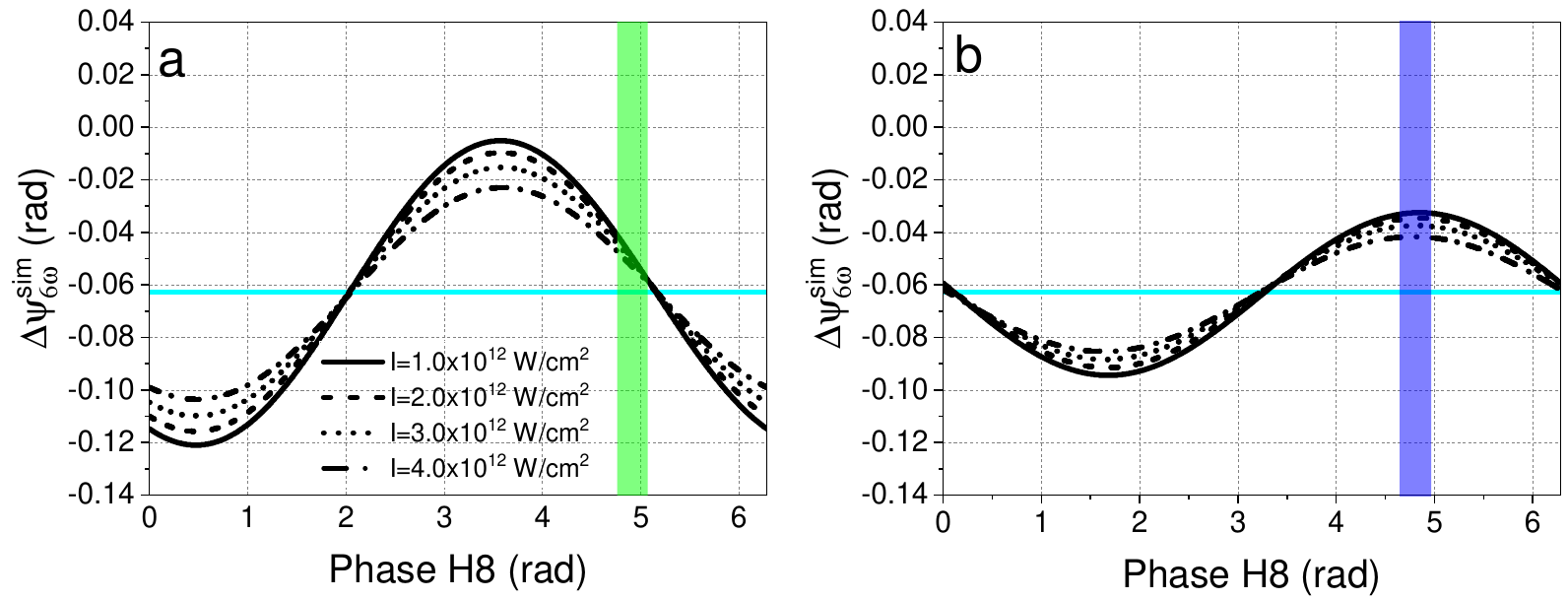}}
\centering \resizebox{1.0\hsize}{!}{\includegraphics{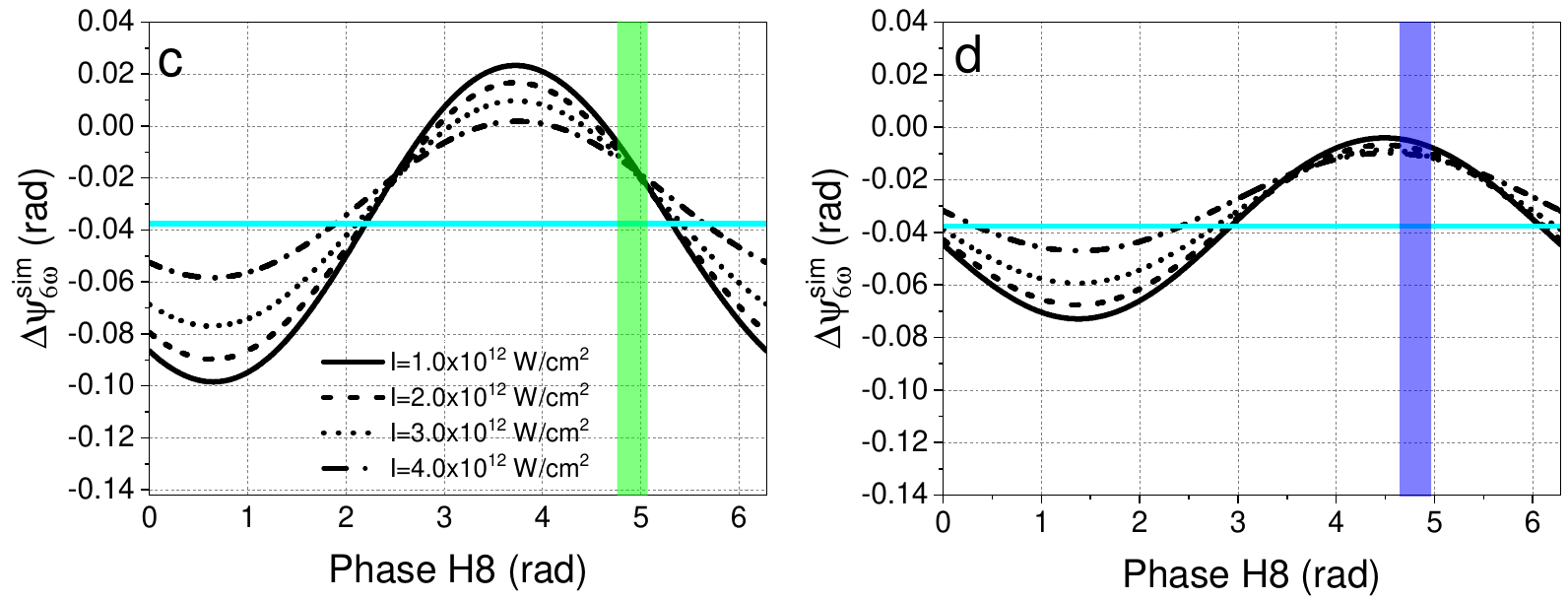}}
\caption{\textbf{Effect of the harmonic $H_8$ on the simulated phase difference $\Delta\Psi^{\mathrm{sim}}_{6\omega}$.} Evolution of the phase $\Delta\Psi^{\mathrm{sim}}_{6\omega}$ for different energy of the emitted photoelectron ($\widetilde{E}_1$, panels a,b; $\widetilde{E}_a$, panels c,d), for the four different intensities  $I=1.0\times10^{12}$~W/cm$^2$, $2.0\times10^{12}$~W/cm$^2$, $3.0\times10^{12}$~W/cm$^2$, and $4.0\times10^{12}$~W/cm$^2$, and for the two second-order phase difference $\Delta\varphi_{6,7,9,10}=\Delta\varphi_A$ (a,c) and $\Delta\varphi_B$ (b,d). The green (blue) shaded area indicate the interval corresponding to the best estimation of the value of the phase of the harmonic $H_8$. The cyan line indicates the value of $\Delta\Psi^{\mathrm{sim}}_{6\omega}$ obtained without the contribution of the harmonic $H_8$ ($F_8=0$) for the energy $\widetilde{E}_1$ (a,b) and $\widetilde{E}_a$ (c,d). Panels a and b: simulations in helium. Panels c and d: simulations in neon.}
\label{Fig3SI}
\end{figure}

The experimental values for the phase differences $\Delta\chi^{\mathrm{exp}}_{3\omega}$, $\Delta\chi^{\mathrm{exp};(+0,0-)}_{6\omega}$, and $\Delta\Psi^{\mathrm{exp}}_{6\omega}$ were obtained as the difference of the corresponding values of $\chi^{\mathrm{exp}}_{3\omega}$ and $\chi^{\mathrm{exp}}_{6\omega}$ obtained by the fit of the experimental data according to Eq.~\ref{Eq1}.
The presence of the harmonic $H_8$ introduces additional pathways contributing to the sidebands between consecutive ($H_6$-$H_7$ and $H_9$-$H_{10}$) and non-consecutive harmonics ($H_7$-$H_9$) characterised by the absorption of one photon of the harmonic $H_8$ and the exchange of a single or multiple NIR photons. The effect of these contributions on the phases of the sideband oscillations needs to be corrected in order to compare the experimental phase shift with the TDSE predictions.
Towards this goal, we have performed TDSE simulations with and without the contribution of the harmonic $H_8$ considering the amplitudes and phases determined from the fitting procedure. We have then extracted from both sets of simulations the oscillations at frequencies $3\omega$ and $6\omega$ for all sidebands and determined for both sets of simulations independently the phase differences $\Delta\chi^{\mathrm{sim}}_{3\omega}$,  $\Delta\chi_{6\omega}^{\mathrm{sim};(+0,0-)}$, and $\Delta\Psi^{\mathrm{sim}}_{3\omega}$.
The results are presented in~\ref{Fig1SI},~\ref{Fig2SI}, and~\ref{Fig3SI} for the phase differences $\Delta\chi^{\mathrm{sim}}_{3\omega}$,  $\Delta\chi^{\mathrm{sim};(+0,0-)}_{6\omega}$, and  $\Delta\Psi^{\mathrm{sim}}_{3\omega}$, respectively.

In the figures, the green and blue shaded areas indicate the estimated values (and the corresponding error bars) for the phase of the harmonic $H_8$. In the figures, we also present the expected value without the presence of the harmonic $H_8$, obtained by setting the amplitude of the latter equal to zero: $F_8=0$.
The correction due to the presence of the harmonic $H_8$ was estimated as the difference between the values obtained with and without the harmonic $H_8$:
\begin{eqnarray}\label{corr}
  \Delta\chi^{\mathrm{corr}}_{3\omega}&=&\Delta\chi^{\mathrm{sim}}_{3\omega}[F_8\neq0]-\Delta\chi^{\mathrm{sim}}_{3\omega}[F_8=0]; \nonumber\\ \Delta\chi^{\mathrm{corr};(+0,0-)}_{6\omega}&=&\Delta\chi^{\mathrm{sim};(+0,0-)}_{6\omega}[F_8\neq0]-\Delta\chi^{\mathrm{sim};(+0,0-)}_{6\omega}[F_8=0]\nonumber\\
  \Delta\Psi^{\mathrm{corr}}_{6\omega}&=&\Delta\Psi^{\mathrm{sim}}_{6\omega}[F_8\neq0]-\Delta\Psi^{\mathrm{sim}}_{6\omega}[F_8=0]
\end{eqnarray}
The error in the estimation of the corrections was determined by considering the (maximum) variation of the phase difference in the interval corresponding to the uncertainty in the estimation of the phase of the harmonic $H_8$ (shaded areas).

These values are then used to correct the experimental values according to the relations:
\begin{equation}\label{EqS1}
 \Delta\chi_{3\omega}=\Delta\chi^{\mathrm{exp}}_{3\omega}-\Delta\chi^{\mathrm{corr}}_{3\omega};\quad\Delta\chi^{(+0,0-)}_{6\omega}=\Delta\chi^{\mathrm{exp};(+0,0-)}_{6\omega}-\Delta\chi^{\mathrm{corr};(+0,0-)}_{6\omega}; \Delta\Psi_{6\omega}=\Delta\Psi^{\mathrm{exp}}_{6\omega}-\Delta\Psi^{\mathrm{corr}}_{6\omega}
\end{equation}
where the values $\Delta\chi^{\mathrm{exp}}_{3\omega}$ , $\Delta\chi^{\mathrm{exp};(+0,0-)}_{6\omega}$ , and $\Delta\Psi^{\mathrm{exp}}_{3\omega}$ are the phase differences extracted directly from the experimental data.

\textcolor{black}{For the simulations, the values of the phase differences $\Delta\chi_{3\omega}$, $\Delta\chi_{6\omega}^{(+0,0-)}$, and $\Delta\Psi_{3\omega}$ presented in Figs.~\ref{Fig3} and \ref{Fig4SI} are equal to the values obtained by the TDSE simulations considering $F_8=0$.}

The values for the experimental phase differences measured in helium, the corresponding corrections (with an additional minus sign), and their sums for the two different second-order phase differences are presented in~\ref{Table4},~\ref{Table5},~\ref{Table6}, and~\ref{Table7}. The correction for the $3\omega$ oscillations can be as high as 750~mrad. Conversely, the corrections for the components oscillating at $6\omega$ are smaller (up to approximately 20~mrad) thus supporting the robustness of the conclusions that can be drawn from the measurements at this oscillation frequency.

\clearpage

\begin{table}[h]
  \centering
\begin{tabular}{| c | c | c | c |}
\hline \multicolumn{4} {|c|} {\bfseries $\Delta\chi_{3\omega}$ \textbf{Helium}}\\ \hline\hline
  $I$ & $\Delta\chi^{\mathrm{exp}}_{3\omega}(\times10^{-2})$ (rad) & $-\Delta\chi^{\mathrm{corr}}_{3\omega}(\times10^{-2})$ (rad) & $\Delta\chi_{3\omega}(\times10^{-2})$ (rad)  \\ \hline
\multicolumn{4} {|c|} {$\Delta\varphi_{6,7,9,10}=\Delta\varphi_A$;\quad $E_1$=6.51 eV;\quad $\Delta\chi^{\mathrm{sim}}_{3\omega}[F_8=0]=-0.0412$ (rad)}\\ \hline
    $I_1$  & $3.59\pm3.58$ & $0.149\pm0.027$     & $3.74\pm3.61$   \\ \hline  %& $0.0173\pm0.209$  & ..  & ..
    $I_2$  & $1.21\pm4.05$ & $-0.401\pm0.050$    & $0.81\pm4.10$    \\ \hline %& $0.0522\pm0.0231$ & ..  & ..
    $I_3$  & $-0.22\pm2.39$ & $-1.69\pm0.11$     & $-1.91\pm2.50$   \\ \hline % & $0.0387\pm0.0398$ & ..  & ..
\multicolumn{4} {|c|} {$\Delta\varphi_{6,7,9,10}=\Delta\varphi_A$;\quad $E_2$=20.50 eV;\quad $\Delta\chi^{\mathrm{sim}}_{3\omega}[F_8=0]=-0.0055$ (rad)}\\ \hline
    $I_1$  & $1.73\pm2.09$ &  $-0.510\pm0.064$     & $1.22\pm2.15$    \\ \hline  %& $0.0173\pm0.209$  & ..  & ..
    $I_2$  & $5.22\pm2.31$ & $-1.245\pm0.143$     & $3.97\pm2.45$     \\ \hline %& $0.0522\pm0.0231$ & ..  & ..
    $I_3$  & $3.87\pm3.98$ & $-3.486\pm0.328$     & $0.38\pm4.31$     \\ \hline\hline %& $0.0387\pm0.0398$ & ..  & ..
    \multicolumn{4} {|c|} {$\Delta\varphi_{6,7,9,10}=\Delta\varphi_B$;\quad $E_1$=6.51 eV;\quad $\Delta\chi^{\mathrm{sim}}_{3\omega}[F_8=0]=-0.0412$ (rad)}\\ \hline
    $I_1$  & $-3.48\pm2.26$ & $0.601\pm0.027$     & $-2.88\pm2.29$   \\ \hline  %& $0.0173\pm0.209$  & ..  & ..
    $I_2$  & $-5.01\pm3.32$ & $0.551\pm0.052$         & $-4.459\pm3.37$    \\ \hline %& $0.0522\pm0.0231$ & ..  & ..
    $I_3$  & $-11.59\pm4.51$ & $0.371\pm0.107$     & $-11.22\pm4.62$    \\ \hline % & $0.0387\pm0.0398$ & ..  & ..
    \multicolumn{4} {|c|} {$\Delta\varphi_{6,7,9,10}=\Delta\varphi_B$;\quad $E_2$=20.50 eV;\quad $\Delta\chi^{\mathrm{sim}}_{3\omega}[F_8=0]=-0.0055$ (rad)}\\ \hline
    $I_1$  & $2.96\pm1.88$ & $-0.52\pm0.081$     & $2.44\pm1.96$   \\ \hline  %& $0.0173\pm0.209$  & ..  & ..
    $I_2$  & $5.34\pm2.09$ & $-1.277\pm0.182$         & $4.074\pm2.272$     \\ \hline %& $0.0522\pm0.0231$ & ..  & ..
    $I_3$  & $-4.00\pm10.23$ & $-3.519\pm0.418$     & $-7.519\pm10.648$     \\ \hline\hline %& $0.0387\pm0.0398$ & ..  & .
\end{tabular}
  \caption{Experimental phase difference $\Delta\chi^{\mathrm{exp}}_{3\omega}$ and corresponding correction with changed sign $-\Delta\chi^{\mathrm{corr}}_{3\omega}$ contributing to the phases $\Delta\chi_{3\omega}$ according to~\ref{EqS1} for the two energies $E_1$ and $E_2$, and for the three different NIR intensities $I_1$, $I_2$, and $I_3$ used in the experiment in helium. The energies $E_1$ and $E_2$ are those of the sidebands $S^{(+)}_{6,7}$ and $S^{(+)}_{9,10}$, respectively. The error bars for the experimental data were obtained by propagation of the errors in the fit of the sideband oscillations. The error bars on the correction values were obtained by propagation of the errors in the estimation of the phase of the harmonic $H_8$. The final error bar on $\Delta\chi_{3\omega}$ was obtained by error propagation according to~\ref{EqS1}.}\label{Table4}
\end{table}

\begin{table}[h]
  \centering
\begin{tabular}{| c | c | c | c |}
\hline \multicolumn{4} {|c|} {\bfseries $\Delta\chi^{(0-)}_{6\omega}$ \textbf{Helium}}\\ \hline\hline
  $I$ & $\Delta\chi^{\mathrm{exp};(0-)}_{6\omega}(\times10^{-2})$ (rad) & $-\Delta\chi^{\mathrm{corr};(0-)}_{6\omega}(\times10^{-2})$ (rad) & $\Delta\chi^{(0-)}_{6\omega}(\times10^{-2})$ (rad)  \\ \hline
\multicolumn{4} {|c|} {$\Delta\varphi_{6,7,9,10}=\Delta\varphi_A$;\quad $\overline{E}_1$=12.73 eV;\quad $\Delta\chi^{\mathrm{sim};(0-)}_{6\omega}[F_8=0]=-0.03421$ (rad)}\\ \hline
    $I_1$  & $-0.7\pm12.4$ & $0.72\pm0.39$     & $0.015\pm12.831$   \\ \hline  %& $0.0173\pm0.209$  & ..  & ..
    $I_2$  & $-9.37\pm9.51$ & $0.72\pm0.36$         & $-8.65\pm9.87$    \\ \hline %& $0.0522\pm0.0231$ & ..  & ..
    $I_3$  & $-6.50\pm4.32$ & $0.79\pm0.27$     & $-5.70\pm4.59$    \\ \hline % & $0.0387\pm0.0398$ & ..  & ..
    \multicolumn{4} {|c|} {$\Delta\varphi_{6,7,9,10}=\Delta\varphi_B$;\quad $\overline{E}_1$=12.73 eV;\quad $\Delta\chi^{\mathrm{sim};(0-)}_{6\omega}[F_8=0]=-0.03421$ (rad)}\\ \hline
    $I_1$  & $-2.13\pm12.24$ & $-1.11\pm0.31$     & $-3.24\pm12.55$   \\ \hline  %& $0.0173\pm0.209$  & ..  & ..
    $I_2$  & $-6.31\pm7.25$ & $-1.04\pm0.29$      & $-7.35\pm7.54$    \\ \hline %& $0.0522\pm0.0231$ & ..  & ..
    $I_3$  & $2.56\pm8.94$ & $-0.72\pm0.27$       & $1.83\pm9.21$    \\ \hline % & $0.0387\pm0.0398$ & ..  & ..
\end{tabular}
  \caption{Experimental phase difference $\Delta\chi^{\mathrm{exp};(0-)}_{6\omega}$ and corresponding correction with changed sign $-\Delta\chi^{\mathrm{corr};(0-)}_{6\omega}$ contributing to the phases $\Delta\chi^{(0-)}_{6\omega}$ according to~\ref{EqS1} for the energy $\overline{E}_1$ and for the three different NIR intensities $I_1$, $I_2$, and $I_3$ used in the experiment in helium. The energy $\overline{E}_1$ is that of the sideband $S^{(0)}_{7,9}$. The error bars for the experimental data were obtained by propagation of the errors in the fit of the sideband oscillations. The error bars on the correction values were obtained by propagation of the errors in the estimation of the phase of the harmonic $H_8$. The final error bar on $\Delta\chi^{(0-)}_{6\omega}$ was obtained by error propagation according to~\ref{EqS1}.}\label{Table5}
\end{table}

\begin{table}[h]
  \centering
\begin{tabular}{| c | c | c | c |}
\hline \multicolumn{4} {|c|} {\bfseries $\Delta\chi^{(+0)}_{6\omega}$ \textbf{Helium}}\\ \hline\hline
  I & $\Delta\chi^{\mathrm{exp};(+0)}_{6\omega}(\times10^{-2})$ (rad) & $-\Delta\chi^{\mathrm{corr};(+0)}_{6\omega}(\times10^{-2})$ (rad) & $\Delta\chi^{(+0)}_{6\omega}(\times10^{-2})$ (rad)  \\ \hline
\multicolumn{4} {|c|} {$\Delta\varphi_{6,7,9,10}=\Delta\varphi_A$;\quad $\overline{E}^*_1$=14.28 eV;\quad $\Delta\chi^{\mathrm{sim};(+0)}_{6\omega}[F_8=0]=-0.02859$ (rad)}\\ \hline
    I$_1$  & $-11.98\pm13.50$ & $-2.06\pm0.48$     & $-14.04\pm13.98$    \\ \hline  %& $0.0173\pm0.209$  & ..  & ..
    I$_2$  & $-4.83\pm7.56$ & $-2.00\pm0.45$         & $-6.83\pm8.01$     \\ \hline %& $0.0522\pm0.0231$ & ..  & ..
    I$_3$  & $-4.42\pm3.55$ & $-1.77\pm0.36$     & $-6.19\pm3.91$    \\ \hline\hline %& $0.0387\pm0.0398$ & ..  & ..
    \multicolumn{4} {|c|} {$\Delta\varphi_{6,7,9,10}=\Delta\varphi_B$;\quad $\overline{E}^*_1$=14.28 eV;\quad $\Delta\chi^{\mathrm{sim};(+0)}_{6\omega}[F_8=0]=-0.02859$ (rad)}\\ \hline
    I$_1$  & $-3.41\pm11.98$ & $-1.91\pm0.29$     & $-5.32\pm0.13$    \\ \hline  %& $0.0173\pm0.209$  & ..  & ..
    I$_2$  & $-6.62\pm5.66$ & $-1.78\pm0.29$         & $-8.40\pm5.95$     \\ \hline %& $0.0522\pm0.0231$ & ..  & ..
    I$_3$  & $-18.80\pm11.34$ & $-1.39\pm0.28$     & $-20.19\pm11.62$    \\ \hline\hline %& $0.0387\pm0.0398$ & ..  & .
\end{tabular}
  \caption{Experimental phase difference $\Delta\chi^{\mathrm{exp};(+0)}_{6\omega}$ and corresponding correction with changed sign $-\Delta\chi^{\mathrm{corr};(+0)}_{6\omega}$ contributing to the phases $\Delta\chi^{(+0)}_{6\omega}$ according to~\ref{EqS1} for the two energies $\overline{E}^*_1$ and for the three different NIR intensities $I_1$, $I_2$, and $I_3$ used in the experiment in helium. The energy $\overline{E}^*_1$ is that of the sideband $S^{(+)}_{7,9}$. The error bars for the experimental data were obtained by propagation of the errors in the fit of the sideband oscillations. The error bars on the correction values were obtained by propagation of the errors in the estimation of the phase of the harmonic $H_8$. The final error bar on $\Delta\chi^{(+0)}_{6\omega}$ was obtained by error propagation according to~\ref{EqS1}.}\label{Table6}
\end{table}

\begin{table}[h]
  \centering
\begin{tabular}{| c | c | c | c |}
\hline \multicolumn{4} {|c|} {\bfseries $\Delta\Psi_{6\omega}$ \textbf{Helium}}\\ \hline\hline
  I & $\Delta\Psi^{\mathrm{exp}}_{6\omega}(\times10^{-2})$ (rad) & $-\Delta\Psi^{\mathrm{corr}}_{6\omega}(\times10^{-2})$ (rad) & $\Delta\Psi_{6\omega}(\times10^{-2})$ (rad)  \\ \hline
\multicolumn{4} {|c|} {$\Delta\varphi_{6,7,9,10}=\Delta\varphi_A$;\quad $\widetilde{E}_1$=14.28 eV;\quad $\Delta\Psi^{\mathrm{sim}}_{6\omega}[F_8=0]=-0.0628$ (rad)}\\ \hline
    I$_1$  & $-12.74\pm17.85$ & $-1.34\pm0.86$     & $-14.08\pm18.71$   \\ \hline  %& $0.0173\pm0.209$  & ..  & ..
    I$_2$  & $-14.20\pm8.69$ & $-1.28\pm0.79$        & $-15.48\pm9.48$    \\ \hline %& $0.0522\pm0.0231$ & ..  & ..
    I$_3$  & $-10.92\pm4.22$ & $-0.98\pm0.60$   & $-11.90\pm4.82$     \\ \hline % & $0.0387\pm0.0398$ & ..  & ..
    \multicolumn{4} {|c|} {$\Delta\varphi_{6,7,9,10}=\Delta\varphi_B$;\quad $\widetilde{E}_1$=14.28 eV;\quad $\Delta\Psi^{\mathrm{sim}}_{6\omega}[F_8=0]=-0.0628$ (rad)}\\ \hline
    I$_1$  & $-5.67\pm13.55$ & $-3.019\pm0.06$     & $-8.69\pm13.61$    \\ \hline  %& $0.0173\pm0.209$  & ..  & ..
    I$_2$  & $-12.94\pm7.10$ & $-2.817\pm0.05$         & $-15.76\pm7.15$    \\ \hline %& $0.0522\pm0.0231$ & ..  & ..
    I$_3$  & $-16.24\pm3.39$ & $-2.113\pm0.03$     & $-18.35\pm3.42$    \\ \hline % & $0.0387\pm0.0398$ & ..  & ..
\end{tabular}
  \caption{Experimental phase difference $\Delta\Psi^{\mathrm{exp}}_{6\omega}$ and corresponding correction with changed sign $-\Delta\Psi^{\mathrm{corr}}_{6\omega}$ contributing to the phases $\Delta\Psi_{6\omega}$ according to~\ref{EqS1} for the energy $\widetilde{E}_1$, and for the three different NIR intensities $I_1$, $I_2$, and $I_3$ used in the experiment in helium. The error bars for the experimental data were obtained by propagation of the errors in the fit of the sideband oscillations. The energy $\widetilde{E}_1$ is that of the sideband $S^{(+)}_{7,9}$. The error bars on the correction values were obtained by propagation of the errors in the estimation of the phase of the harmonic $H_8$. The final error bar on $\Delta\Psi_{6\omega}$ was obtained by error propagation according to~\ref{EqS1}.}\label{Table7}
\end{table}
\clearpage

\subsection*{Experimental data and theoretical simulations in neon}
\begin{figure}[ht]
\centering \resizebox{1.0\hsize}{!}{\includegraphics{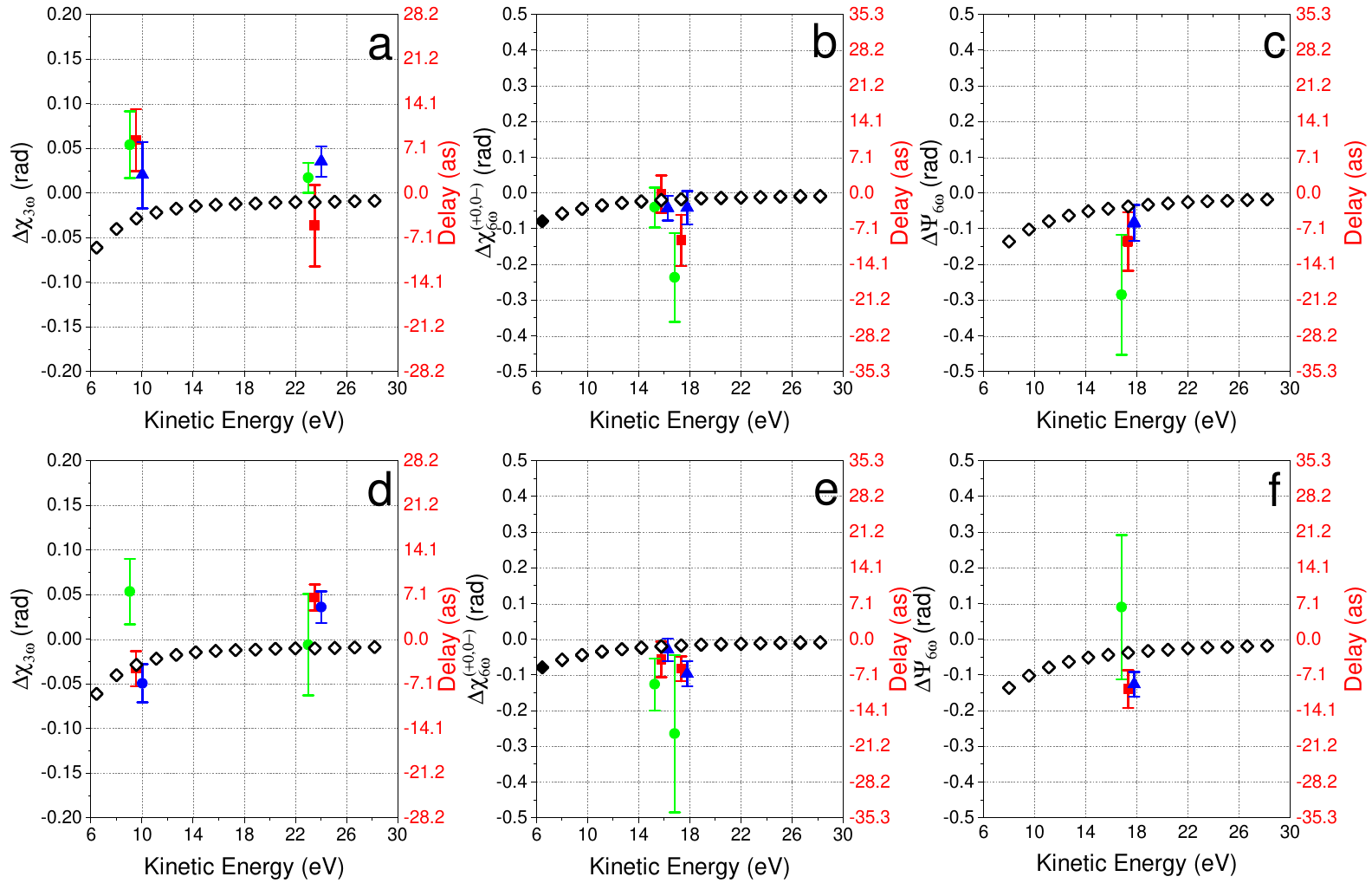}}
\caption{\textbf{Comparison between simulations and experimental results for the phase differences $\Delta\chi_{3\omega}$, $\Delta\chi^{(+0,0-)}_{6\omega}$, and $\Delta\Psi_{6\omega}$ in neon.} Comparison of the theoretical simulations (diamonds) obtained using the TDSE and the experimental data (symbols) for the phase differences $\Delta\chi_{3\omega}$ (a,d), $\Delta\chi^{(+0,0-)}_{6\omega}$ (b,e), and $\Delta\Psi_{6\omega}$ (c,f) for three different NIR intensities: $I_a=1.2\times10^{12}$~W/cm$^2$ (green circles), $I_b=2.2\times10^{12}$~W/cm$^2$ (red squares), and $I_c=3.0\times10^{12}$~W/cm$^2$ (blue triangles). The second-order phase difference $\Delta\varphi_{6,7,9,10}$ for the measurements presented in panels (a-c) and (d-f) were $\Delta\varphi_A$ and $\Delta\varphi_B$, respectively. The experimental data were corrected for the contribution of the harmonic $H_8$. The error bars are derived by propagation of the uncertainties in the fitting procedure of the sideband oscillations. The open (full) diamonds in panels (b,e) indicate the phase difference $\Delta\chi_{6\omega}^{(+0)}$ ($\Delta\chi_{6\omega}^{(0-)}$).}
\label{Fig4SI}
\end{figure}

\ref{Fig4SI} presents the comparison between the experimental data measured in neon with the theoretical simulations performed using the single-active electron model with the model potential from Ref.~\citenum{tongEmpiricalFormulaStatic2005}.
The data were acquired for three different NIR intensities: $I_a=1.2\times10^{12}$~W/cm$^2$, $I_b=2.2\times10^{12}$~W/cm$^2$, $I_c=3.0\times10^{12}$~W/cm$^2$.
The effect of the harmonic $H_8$ on the phase differences of the sidebands was corrected for the neon case. The values of the phases and the corresponding corrections and final results for the neon-case are presented in~\ref{Table8},~\ref{Table9},~\ref{Table10}, and~\ref{Table11}.

\begin{table}[h]
  \centering
\begin{tabular}{| c | c | c | c |}
\hline \multicolumn{4} {|c|} {\bfseries $\Delta\chi_{3\omega}$ \textbf{Neon}}\\ \hline\hline
  I & $\Delta\chi^{\mathrm{exp}}_{3\omega}(\times10^{-2})$ (rad) & $-\Delta\chi^{\mathrm{corr}}_{3\omega}(\times10^{-2})$ (rad) & $\Delta\chi_{3\omega}(\times10^{-2})$ (rad)  \\ \hline
\multicolumn{4} {|c|} {$\Delta\varphi_{6,7,9,10}=\Delta\varphi_A$;\quad $E_a$=9.53 eV;\quad $\Delta\chi^{\mathrm{sim}}_{3\omega}[F_8=0]=-0.02781$ (rad)}\\ \hline
    I$_a$  & $5.86\pm3.70$     & $-0.48\pm0.06.1$     & $5.38\pm3.75$   \\ \hline  %& $0.0173\pm0.209$  & ..  & ..
    I$_b$  & $7.14\pm3.32$          & $-1.20\pm0.13$    & $5.942\pm3.45$    \\ \hline %& $0.0522\pm0.0231$ & ..  & ..
    I$_c$  & $3.64\pm3.51$         & $-1.62\pm0.22$     & $2.017\pm3.73$   \\ \hline % & $0.0387\pm0.0398$ & ..  & ..
\multicolumn{4} {|c|} {$\Delta\varphi_{6,7,9,10}=\Delta\varphi_A$;\quad $E_b$=23.53 eV;\quad $\Delta\chi^{\mathrm{sim}}_{3\omega}[F_8=0]=-0.00995$ (rad)}\\ \hline
    I$_a$  & $2.47\pm1.63$ &  $-0.75\pm0.07$     & $1.75\pm1.70$    \\ \hline  %& $0.0173\pm0.209$  & ..  & ..
    I$_b$  & $-2.13\pm4.38$ & $-1.53\pm0.18$     & $-3.66\pm4.56$     \\ \hline %& $0.0522\pm0.0231$ & ..  & ..
    I$_c$  & $6.51\pm1.35$ & $-3.00\pm0.33$     & $3.52\pm1.68$     \\ \hline\hline %& $0.0387\pm0.0398$ & ..  & ..
    \multicolumn{4} {|c|} {$\Delta\varphi_{6,7,9,10}=\Delta\varphi_B$;\quad $E_a$=9.53 eV;\quad $\Delta\chi^{\mathrm{sim}}_{3\omega}[F_8=0]=-0.02781$ (rad)}\\ \hline
    I$_a$  & $4.80\pm3.57$ & $0.55\pm0.062$     & $5.35\pm3.63$   \\ \hline  %& $0.0173\pm0.209$  & ..  & ..
    I$_b$  & $-4.26\pm1.82$ & $0.991\pm0.14$         & $-3.27\pm1.96$    \\ \hline %& $0.0522\pm0.0231$ & ..  & ..
    I$_c$  & $-6.81\pm1.91$ & $1.843\pm0.23$     & $-4.97\pm2.14$    \\ \hline % & $0.0387\pm0.0398$ & ..  & ..
    \multicolumn{4} {|c|} {$\Delta\varphi_{6,7,9,10}=\Delta\varphi_B$;\quad $E_b$=23.53 eV;\quad $\Delta\chi^{\mathrm{sim}}_{3\omega}[F_8=0]=-0.00995$ (rad)}\\ \hline
    I$_a$  & $0.12\pm5.59$ & $-0.761\pm0.091$     & $-0.64\pm5.68$   \\ \hline  %& $0.0173\pm0.209$  & ..  & ..
    I$_b$  & $6.22\pm1.27$ & $-1.54\pm0.22$         & $4.68\pm1.49$     \\ \hline %& $0.0522\pm0.0231$ & ..  & ..
    I$_c$  & $6.61\pm1.36$ & $-3.00\pm0.41$     & $3.61\pm1.77$     \\ \hline\hline %& $0.0387\pm0.0398$ & ..  & .
\end{tabular}
  \caption{Experimental phase difference $\Delta\chi^{\mathrm{exp}}_{3\omega}$ and corresponding correction with changed sign $-\Delta\chi^{\mathrm{corr}}_{3\omega}$ contributing to the phases $\Delta\chi_{3\omega}$ according to~\ref{EqS1} for the two energies $E_a$ and $E_b$, and for the three different NIR intensities $I_a$, $I_b$, and $I_c$ used in the experiment in neon. The energies $E_a$ and $E_b$ are those of the sidebands $S^{(+)}_{6,7}$ and $S^{(+)}_{9,10}$, respectively. The error bars for the experimental data were obtained by propagation of the errors in the fit of the sideband oscillations. The error bars on the correction values were obtained by propagation of the errors in the estimation of the phase of the harmonic $H_8$. The final error bar on $\Delta\chi_{3\omega}$ was obtained by error propagation according to~\ref{EqS1}.}\label{Table8}
\end{table}

\begin{table}[h]
  \centering
\begin{tabular}{| c | c | c | c |}
\hline \multicolumn{4} {|c|} {\bfseries $\Delta\chi^{(0-)}_{6\omega}$ \textbf{Neon}}\\ \hline\hline
  I & $\Delta\chi^{\mathrm{exp};(0-)}_{6\omega}(\times10^{-2})$ (rad) & $-\Delta\chi^{\mathrm{corr};(0-)}_{6\omega}(\times10^{-2})$ (rad) & $\Delta\chi^{(0-)}_{6\omega}(\times10^{-2})$ (rad)  \\ \hline
\multicolumn{4} {|c|} {$\Delta\varphi_{6,7,9,10}=\Delta\varphi_A$;\quad $\overline{E}_a$=15.76 eV;\quad $\Delta\chi^{\mathrm{sim};(0-)}_{6\omega}[F_8=0]=-0.02058$ (rad)}\\ \hline
    I$_a$  & $-4.34\pm5.28$ & $0.29\pm0.30$     & $-4.05\pm5.58$   \\ \hline  %& $0.0173\pm0.209$  & ..  & ..
    I$_b$  & $-0.76\pm5.04$ & $0.44\pm0.25$     & $-0.32\pm5.29$    \\ \hline %& $0.0522\pm0.0231$ & ..  & ..
    I$_c$  & $-4.86\pm3.29$ & $0.65\pm0.19$     & $-4.21\pm3.48$    \\ \hline % & $0.0387\pm0.0398$ & ..  & ..
    \multicolumn{4} {|c|} {$\Delta\varphi_{6,7,9,10}=\Delta\varphi_B$;\quad $\overline{E}_a$=15.76 eV;\quad $\Delta\chi^{\mathrm{sim};(0-)}_{6\omega}[F_8=0]=-0.02058$ (rad)}\\ \hline
    I$_a$  & $-11.73\pm7.14$ & $-0.92\pm0.20$     & $-12.65\pm7.34$   \\ \hline  %& $0.0173\pm0.209$  & ..  & ..
    I$_b$  & $-5.01\pm4.77$ & $-0.72\pm0.21$     & $-5.73\pm4.98$    \\ \hline %& $0.0522\pm0.0231$ & ..  & ..
    I$_c$  & $-2.50\pm2.99$ & $-0.46\pm0.20$     & $-2.96\pm3.19$    \\ \hline % & $0.0387\pm0.0398$ & ..  & ..
\end{tabular}
  \caption{Experimental phase difference $\Delta\chi^{\mathrm{exp};(0-)}_{6\omega}$ and corresponding correction with changed sign $-\Delta\chi^{\mathrm{corr};(0-)}_{6\omega}$ contributing to the phases $\Delta\chi^{(0-)}_{6\omega}$ according to~\ref{EqS1} for the energy $\overline{E}_a$ and for the three different NIR intensities $I_a$, $I_b$, and $I_c$ used in the experiment in neon. The energy $\overline{E}_a$ is that of the sideband $S^{(0)}_{7,9}$. The error bars for the experimental data were obtained by propagation of the errors in the fit of the sideband oscillations. The error bars on the correction values were obtained by propagation of the errors in the estimation of the phase of the harmonic $H_8$. The final error bar on $\Delta\chi^{(0-)}_{6\omega}$ was obtained by error propagation according to~\ref{EqS1}.}\label{Table9}
\end{table}

\begin{table}[h]
  \centering
\begin{tabular}{| c | c | c | c |}
\hline \multicolumn{4} {|c|} {\bfseries $\Delta\chi^{(+0)}_{6\omega}$ \textbf{Neon}}\\ \hline\hline
  I & $\Delta\chi^{\mathrm{exp};(+0)}_{6\omega}(\times10^{-2})$ (rad) & $-\Delta\chi^{\mathrm{corr};(+0)}_{6\omega}(\times10^{-2})$ (rad) & $\Delta\chi^{(+0)}_{6\omega}(\times10^{-2})$ (rad)  \\ \hline
\multicolumn{4} {|c|} {$\Delta\varphi_{6,7,9,10}=\Delta\varphi_A$;\quad $\overline{E}^*_a$=17.32 eV;\quad $\Delta\chi^{\mathrm{sim};(+0)}_{6\omega}[F_8=0]=-0.01699$ (rad)}\\ \hline
    I$_a$  & $-21.06\pm11.78$ & $-2.61\pm0.59$     & $-23.66\pm12.37$    \\ \hline  %& $0.0173\pm0.209$  & ..  & ..
    I$_b$  & $-10.63\pm6.63$ & $-2.59\pm0.53$     & $-13.22\pm7.16$     \\ \hline %& $0.0522\pm0.0231$ & ..  & ..
    I$_c$  & $-1.40\pm4.22$ & $-2.72\pm0.45$     & $-4.12\pm4.67$   \\ \hline\hline %& $0.0387\pm0.0398$ & ..  & ..
    \multicolumn{4} {|c|} {$\Delta\varphi_{6,7,9,10}=\Delta\varphi_B$;\quad $\overline{E}^*_a$=17.32 eV;\quad $\Delta\chi^{\mathrm{sim};(+0)}_{6\omega}[F_8=0]=-0.01699$ (rad)}\\ \hline
    I$_a$  & $-24.16\pm21.72$ & $-2.27\pm0.38$     & $-26.433\pm22.10$    \\ \hline  %& $0.0173\pm0.209$  & ..  & ..
    I$_b$  & $-6.04\pm3.09$ & $-2.21\pm0.37$     & $-8.25\pm3.46$    \\ \hline %& $0.0522\pm0.0231$ & ..  & ..
    I$_c$  & $-7.38\pm3.13$ & $-2.30\pm0.34$     & $-9.68\pm3.47$    \\ \hline\hline %& $0.0387\pm0.0398$ & ..  & .
\end{tabular}
  \caption{Experimental phase difference $\Delta\chi^{\mathrm{exp};(+0)}_{6\omega}$ and corresponding correction with changed sign $-\Delta\chi^{\mathrm{corr};(+0)}_{6\omega}$ contributing to the phases $\Delta\chi^{(+0)}_{6\omega}$ according to~\ref{EqS1} for the energy $\overline{E}^*_a$ and for the three different NIR intensities $I_a$, $I_b$, and $I_c$ used in the experiment in neon. The energy $\widetilde{E}^*_a$ is that of the sideband $S^{(+)}_{7,9}$. The error bars for the experimental data were obtained by propagation of the errors in the fit of the sideband oscillations. The error bars on the correction values were obtained by propagation of the errors in the estimation of the phase of the harmonic $H_8$. The final error bar on $\Delta\chi^{(+0)}_{6\omega}$ was obtained by error propagation according to~\ref{EqS1}.}\label{Table10}
\end{table}

\begin{table}[h]
  \centering
\begin{tabular}{| c | c | c | c |}
\hline \multicolumn{4} {|c|} {\bfseries $\Delta\Psi_{6\omega}$ \textbf{Neon}}\\ \hline\hline
  I & $\Delta\Psi^{\mathrm{exp}}_{6\omega}(\times10^{-2})$ (rad) & $-\Delta\Psi^{\mathrm{corr}}_{6\omega}(\times10^{-2})$ (rad) & $\Delta\Psi_{6\omega}(\times10^{-2})$ (rad)  \\ \hline
\multicolumn{4} {|c|} {$\Delta\varphi_{6,7,9,10}=\Delta\varphi_A$;\quad $\widetilde{E}_a$=17.32 eV;\quad $\Delta\Psi^{\mathrm{sim}}_{6\omega}[F_8=0]=-0.03757$ (rad)}\\ \hline
    I$_a$  & $-26.16\pm16.00$ & $-2.32\pm0.88$     & $-28.47\pm16.88$   \\ \hline  %& $0.0173\pm0.209$  & ..  & ..
    I$_b$  & $-11.41\pm7.44$ & $-2.14\pm0.78$     & $-13.55\pm8.22$    \\ \hline %& $0.0522\pm0.0231$ & ..  & ..
    I$_c$  & $-6.30\pm4.42$ & $-2.08\pm0.63$     & $-8.38\pm5.05$     \\ \hline % & $0.0387\pm0.0398$ & ..  & ..
    \multicolumn{4} {|c|} {$\Delta\varphi_{6,7,9,10}=\Delta\varphi_B$;\quad $\widetilde{E}_a$=17.32 eV;\quad $\Delta\Psi^{\mathrm{sim}}_{6\omega}[F_8=0]=-0.03757$ (rad)}\\ \hline
    I$_a$  & $12.17\pm19.91$ & $-3.19\pm0.20$     & $8.98\pm20.11$    \\ \hline  %& $0.0173\pm0.209$  & ..  & ..
    I$_b$  & $-11.05\pm5.18$ & $-2.94\pm0.18$     & $-13.99\pm5.36$    \\ \hline %& $0.0522\pm0.0231$ & ..  & ..
    I$_c$  & $-9.88\pm3.33$ & $-2.77\pm0.15$     & $-12.65\pm3.48$    \\ \hline % & $0.0387\pm0.0398$ & ..  & ..
\end{tabular}
  \caption{Experimental phase difference $\Delta\Psi^{\mathrm{exp}}_{6\omega}$ and correction with changed sign $-\Delta\Psi^{\mathrm{corr}}_{6\omega}$ contributing to the phases $\Delta\Psi_{6\omega}$ according to~\ref{EqS1} for the energy $\widetilde{E}_a$, and for the three different NIR intensities $I_a$, $I_b$, and $I_c$ used in the experiment in neon. The energy $\widetilde{E}_a$ is that of the sideband $S^{(+)}_{7,9}$. The error bars for the experimental data were obtained by propagation of the errors in the fit of the sideband oscillations. The error bars on the correction values were obtained by propagation of the errors in the estimation of the phase of the harmonic $H_8$. The final error bar on $\Delta\Psi_{6\omega}$ was obtained by error propagation according to~\ref{EqS1}.}\label{Table11}
\end{table}

\clearpage

\end{document}